# Policy-Governed RAG — Research Design Study

## Cryptographic Receipts for Audit-Ready Generation

*Version 1.0 — October 2025*

*Prepared for controlled evaluation in compliance- and audit-bound domains.*

**Jean-Marie Le Ray**

Independent researcher, Rome, Italy

jmleray@translation2.com

## ABSTRACT

A policy-governed RAG architecture is specified for audit-ready generation in regulated workflows, organized as a triptych: (I) Contracts/Control (SHRDLU-like), which governs output adherence to legal and internal policies; (II) Manifests/Trails (Memex-like), which cryptographically anchors all cited source evidence to ensure verifiable provenance; and (III) Receipts/Verification (Xanadu-like), which provides the final, portable proof of compliance for auditors (portable COSE/JOSE) (see §4 and Appendix A). Rather than explaining model internals, outputs are gated ex-ante and bound to cryptographically verifiable evidence for each material answer. Unvalidated targets are stated (≥20% relative reduction in confident errors; p95 latency ≤ 900 ms; ≤ 2.2× serve cost) together with a pre-registered (optional) pilot using NO-GO gates. The design complements existing RAG/guardrails by making policy checks auditable, replayable, and receipt-backed. Target domains include back-office compliance in pharma, medical devices, finance, legal, and the public sector where error costs mays exceed thousands of euros and audit trails are mandatory under regulations such as the EU AI Act. Future evaluations may pre-commit to publishing negative results when any example NO-GO gate is not met.

## TABLE OF CONTENTS

# GLOSSARY

## Table of Acronyms & Terms

Acronyms and specialized terms used in this document are defined in the Glossary.
The first occurrence of each term is not specially marked.

| Acronym / Term | Expansion | Brief meaning in this document |
|---|---|---|
| ANN | Approximate Nearest Neighbor | Indexing/search method used in dense retrieval. |
| ASR | Attack Success Rate | Share of successful red-team attacks. |
| BY | Benjamini–Yekutieli | FDR control under arbitrary dependence (used for supports). |
| confident-error@τ | Primary endpoint for safe emission | Share of emitted answers that are wrong despite claimed confidence ≥ τ; tracked per route and used for power planning and gating. |
| COSE | CBOR Object Signing and Encryption | Portable binary signature/encryption format for receipts. |
| DPR | Dense Passage Retrieval | Dense retriever commonly used in RAG pipelines. |
| ECDSA | Elliptic-Curve Digital Signature Algorithm | Digital signature algorithm used for receipts/anchors. |
| FDR | False Discovery Rate | Expected share of accepted supports that are actually false. |
| FWER / FWE | Family-Wise Error Rate | Probability of ≥1 false positive across multiple tests. |
| G_indep | Graph-independence score | Measure of citation diversity; gate typically ≥ 0.70. |
| GDPR | General Data Protection Regulation (EU) | EU data protection and privacy regulation; governs PII/PHI handling and disclosure. |
| Holm | Holm (1979) procedure | Step-down correction controlling FWER (used for contradictions). |
| HSM | Hardware Security Module | Secure hardware for key storage and signing. |
| ICL | In-Context Learning (score) | Signal/score used to trigger heavy verification under adversarial indicators. |
| JOSE | JSON Object Signing and Encryption | JSON-based signature/encryption format for receipts. |
| Justification threshold ($J \geq \tau_J$) | Decision threshold for emission | Configured cutoff on the composite justification score J; emission allowed only if $J \geq \tau_J$ under multiplicity control (FDR/FWER). |
| KID | Key Identifier | Identifier for signing keys included in receipts. |
| KRN | Key Revocation Notice | Registry tracking key validity and revocations. |
| Lite Receipt | (Term of art) | Minimal portable receipt with scope diagnostics, proofs, and signature only. |
| m_eff | Effective number of tests | Estimate of independent checks under dependence; used for multiplicity control. |
| MDCG | Medical Device Coordination Group (EU) | EU body issuing guidance cited in domain examples. |
| Memex-like | The Recorder | Research trails/provenance metaphor for Manifests/Trails. |
| MSES | Minimal Sufficient Evidence Set | Smallest evidence set that still passes policy checks. |
| MiFID II | Markets in Financial Instruments Directive II (EU) | EU rules for financial markets and investor protection; relevant for compliance workflows. |
| NO-GO | Go/No-Go gate | Pre-registered (optional) threshold that blocks progression if unmet. |
| PII / PHI | Personally Identifiable / Protected Health Information | Sensitive data types; trigger privacy gates/redaction. |
| Promotion Receipt | Signed snapshot-promotion record | Signed artifact emitted at deployment promotion capturing snapshot IDs (SHA-256), |
| | | policy/route versions, thresholds, verifier outcomes, and SLO checks. |
| Q2S | Queries-to-Success | Average number of queries needed to achieve a successful attack. |
| RAG | Retrieval-Augmented Generation | LLM answers grounded in retrieved sources. |
| SHA-256 | Secure Hash Algorithm 256-bit | Cryptographic hash (FIPS 180-4) producing a 256-bit digest for content addressing and integrity checks. Identical bytes → same hash; any change → a different digest. |
| SHRDLU-2025/Memex-2025/Xanadu-2025 Triptych | Triptych revived by AI in 2025. Historic labels are mnemonics only, not product designations. | Each historic project provides a modern governance function for Large Language Models (LLMs) |
| SHRDLU-like | The Planner | Rules-first control/gating metaphor for Contracts/Control. |
| SLAs / SLOs | Service-Level Agreements / Objectives | Performance & reliability targets (latency, proof size, etc.). |
| SMT | Sparse Merkle Tree | Merkle tree over a fixed keyspace with hash-addressed leaves; supports inclusion and non-inclusion proofs and compact multiproofs. |
| SOX | Sarbanes–Oxley Act (US) | US corporate accountability and audit controls; impacts record-keeping and attestation. |
| TTD | Time-to-Detect | Elapsed time to detect an attack or anomaly. |
| TTL | Time-to-Live | Freshness/validity window for data or manifests. |
| Xanadu-like | The Auditor | Signed receipts/verification metaphor for Receipts/Verification. |

## DISCLAIMER

I am not an AI researcher; my background is in machine translation. This article presents a pre-pilot technical specification for policy-compliant RAG in regulated domains. It was developed in a Human–LLMs collaborative authorship process with assistance from ChatGPT (GPT-5 Thinking), Claude (Sonnet 4.5), and Grok 4 Fast; the author retains full responsibility for the final text. Feedback is welcome.

The specification adapts three governance primitives—contracts, trails, receipts—to modern LLM workflows (mnemonics: SHRDLU control, Memex provenance, Xanadu verification). It provides (i) an architecture and minimal artifacts (schemas, verifiers, challenge sets), (ii) evaluation protocols using MSES, selective prediction, and Holm–Bonferroni/BY multiplicity control, and (iii) signed receipts with cryptographic proofs.

Non-claims. This is not a deployed or validated system and does not offer empirical ROI, production experience, truth guarantees, or mechanistic interpretability. Performance figures are targets, not measurements (e.g., projected 15–35% reduction in confident errors on hard-fact routes; 20–35% abstention with helpful-abstain UX; overheads ≈2–5× ingest, 1.8–3.2× serve).

Scope. Intended for back-office compliance in heavily regulated sectors (finance, healthcare, legal, public sector) where audit obligations and expected liability costs justify governance overhead. Real-time, creative, or low-stakes uses are out of scope. Progress requires a single-domain pilot (e.g., EU medical-device regulation Q&A) with measured outcomes, user studies, and adversarial testing prior to any operational deployment.

This document is a thought experiment on retrieval-augmented generation (RAG) governance prepared for academic and exploratory research. It does not represent a commercial product specification, grant proposal, or operational commitment. Implementation or deployment decisions remain outside the scope of this paper.

### VALIDATION STATUS

• No production deployments

• No user studies or A/B tests

• Every performance number (latency/overhead/error) is non-binding engineering guidance until D2 passes

• Statistical controls (MSES, Holm, BY, selective prediction) are untested in production

• Cost figures are engineering projections

## LIST OF FIGURES

## EXECUTIVE SUMMARY

Problem. Regulated workflows require auditable answers, not narratives rationalized after the fact. A single wrong answer can be high-impact (e.g., multi-€M penalties). Typical RAG stacks cannot demonstrate what was checked, which sources were in-policy, or why an answer was allowed to ship.

Approach (research design). The design enforces execution contracts at generation time; anchors citations in Merkle-backed manifests; and issues signed receipts that disclose the Minimal Sufficient Evidence Set (MSES), counterfactual flips, and escalation records. Receipts are portable (COSE/JOSE), offline-verifiable, and support internal-only operation (on-prem HSM), with optional public transparency explored in later phases.

Non-claims. No guarantees are made regarding source truth or mechanistic interpretability, and not every adversary is assumed to be caught. The contribution is process attestation—evidence of what was checked—not correctness guarantees.

Evaluation thresholds (proposed, non-binding).

- Effectiveness: ≥ 20% relative reduction in confident-error@τ (95% CI).
- Usefulness: ≥ 70% helpful-abstain and ≤ +10% task abandonment vs. control.
- Performance: p95 ≤ 900 ms and ≤ 2.2× serve overhead (with caches).
- Dependence control: m_eff ≥ 1.3 (95% CI lower bound ≥ 1.1).

Fit. Appropriate for back-office compliance (e.g., pharma, medical devices, public sector, finance/legal) where error costs[*] > €10K, audit trails are mandatory, and queries are hard-fact with jurisdiction/date constraints. Not appropriate for real-time trading, consumer chat, creative/ideation, or domains with error costs < €5K.

Adoption path (research profiles).

- Phase A – Core (3-6 mo): manifests + single verifier + Lite Receipts → D1 ops (latency/cost/abstentions; no efficacy claims).

[*] Cost figures are engineering estimates and subject to empirical validation.

- Phase B — Plus (6–12 mo): MSES, G_indep, cascades → D2 efficacy (powered A/B, n ≥ 1,700 per arm).
- Phase C — Full (optional): transparency logs and public verifiability.

Economics (to be validated). If estimated error cost L < €5K, conduct ops-only validation and publish negative-ROI curves (Appendix F). For all evaluations, report P50/P90 serve cost under normal operation and a stress scenario (no cache, heavy=40%), with component breakdowns (see §G.5, Appendix F). Appendix F includes a breakeven calculator and worked examples.

Future directions. Decisions on evaluation design, pre-registration, partner selection, and deployment are intentionally left to future researchers and practitioners who may explore, adapt, or discard this design as appropriate.

*

# LEVEL 1 — CONCEPTS (for decision-makers)

## 1. Introduction: From post-hoc to attestation-by-design

The approach is grounded in retrieval-augmented generation (RAG) [11] and calibrated risk bounds [8]. Each material answer —i.e., any output that could plausibly affect a regulated obligation, external filing, legal position, or monetary exposure ≥ €10K, or that contains PII/PHI or other policy-restricted data—is gated by explicit policy and tied to named sources, with Merkle-tree inclusion proofs [21] and a receipt indicating counterfactuals that would flip the decision. Scope is hard-fact Q&A over allow-listed corpora with explicit jurisdiction and effective-date constraints. Emission requires at least two independent supports, no detected contradictions, in-policy licensing, and calibrated risk bounds [8], with supports FDR-controlled (Benjamini–Yekutieli) and contradictions FWER-controlled (Holm) [2, 1]. When conditions are not met, the system abstains under selective prediction [9, 10] and provides an explanation (Appendix E — Helpful-abstain UX). All checks are recorded as signed artifacts and can be replayed offline by an independent verifier. Selective prediction and multiple-testing controls govern confident emission and abstention so that risk is managed rather than assumed [1, 2] [9, 10]. This approach fits back-office compliance contexts where mistakes are costly and audit trails are mandatory [25, 26, 27, 28]. No claim is made about the truth of sources or about mechanistic interpretability of the model, and real-time trading, creative/ideation, and bedside clinical use—where latency or irreducible ambiguity dominates—are out of scope.

### 1.1 What is claimed (and what is not)

The claim is decision-level attestation: answers are emitted only once enforceable execution contracts are satisfied, supporting evidence is fixed and reproducible, and a signed receipt binds the decision to its policy checks and proofs. Each receipt specifies the minimal sufficient evidence set, applicable jurisdiction and dates, verifier outcomes and risk bounds, and counterfactuals that would flip the decision, allowing an independent party to replay and audit the result offline. This is a commitment to process accountability—what was checked, under which thresholds, with which artifacts—not to model introspection. No assertion is made that cited sources are true in an absolute sense, nor that adversaries can never slip through; provenance and verification bound risk and make failures inspectable, they do not guarantee truth. Universal applicability is not promised: the method targets governed, hard-fact workflows on allow-listed corpora, not creative generation or open-ended domains where such contracts are ill-posed.

### 1.2 Proposed evaluation metrics & example NO-GO gates (informative)

The following targets illustrate how this design could be evaluated. They are non-binding templates to be adapted—or discarded—by future researchers.

- Effectiveness (example). ≥ 20% relative drop in confident-error@τ (95% CI).
- Usefulness (example). Helpful-abstain ≥ 70% (median of 3 raters; 1–5 Likert; ≥4 counts as helpful) and task abandonment ≤ +10% vs. control (95% CI). Example NO-GO: if either fails, record non-efficacy.
- Latency (example). p95 ≤ 900 ms on interactive policy-governed routes[*].
- Cost reporting (example). Report P50/P90 serve cost vs. a strong RAG control and a stress scenario (no cache, heavy = 40%) per §G.5. Target: ≤ 2.2× at P50 under normal caching; stress numbers are reported, not gated. See Appendix F (ROI).
- Risk control (example). Holm (FWER) for contradictions; Benjamini–Yekutieli (FDR) for supports; report m_eff (effective tests) with CIs.
- Dependence floor (example). m_eff ≥ 1.3 (bootstrap 95% CI lower bound ≥ 1.1). If m_eff < 1.5, consider enforcing a data-diversification policy and demoting the route until improved.
- Sample size (illustrative). Power analysis may target 80–90% power to detect a 20% relative drop (two-sided $\alpha = 0.05$ post-multiplicity). Typical baselines imply ~900–3,200 queries per arm (for $p_0 \in [0.30, 0.10]$). If pre-registration is pursued, exact n and power curves should be documented.
- Domain fit (example NO-GO). If estimated error cost L < €5,000, treat confirmatory efficacy as out of scope; restrict to ops-only reporting and document negative ROI (Appendix F).
- Time box (example). If confirmatory n cannot be met within 12 months, publish D1 (ops report) and record a no-go for efficacy in this domain; consider domain shift.

### 1.3 Adoption tiers (staged)

Principle. Schemas are stable; engines are swappable. Interfaces (contracts, manifests, receipts) stay backward-compatible; models/retrievers/verifiers can change if they meet the same schemas and SLOs

#### *Phase A — Core (minimum viable governance)*

Scope: verifiable inputs → evidence → receipt with minimal parts.
Includes: policy contracts + allow-listed retrieval; shard-level provenance manifests; small calibrated verifier; Lite Receipts (selective disclosure, COSE/JOSE) with scope diagnostics + inclusion multiproofs.

---

[*] User-facing routes with governance enabled (retrieval allow-lists, verifier checks, receipts).

Ops: p95 ≤ 900 ms; abstain on margin; HSM sidecar signing/KRN checks.
Not included: MSES, G_indep, heavy verifiers, public anchoring.
Exit: stable confident-error@τ on pilot traffic; proof SLOs met; Lite Receipts accepted by internal audit.

### *Phase B — Plus (evidence rigor & diversity)*

Adds: MSES (greedy elimination); G_indep (K=3, ≤50% per issuer); cascaded verifiers (cheap→small→heavy); multiplicity control (Holm for contradictions, BY for supports).
Ops: temporal diversity on drifting topics; receipts carry MSES, G_indep, adjusted p-values, risk budgets; canaries + shadow eval.
Exit: red-team pass (ASR↓ ≥30%, Q2S↑ ≥3×, p95 TTD ≤15 min) without unacceptable cost (≤2.2× P50).

### *Phase C — Full (transparency & autonomic ops)*

Adds: transparency-log anchoring (CT/Sigstore/Rekor-like); auto-degradation on proof/verify/KRN faults; drift dashboards.
Ops: dual-channel revocation (substrate + local KRN), monthly SLOs, scoped disclosure (internal/partner/public).
Exit: audit-conformant; reproducible replays; incident response within policy SLAs.
Upgrade path: no schema churn; engine swaps must meet calibration/accuracy under the same selective-prediction protocol; promotion/rollback via signed Promotion Receipts.

## 1.4 Regulatory risk classification (informative)

This section provides an informative cross-walk to major regulatory frameworks; it does not constitute a formal classification or compliance claim. Actual risk tier and regulatory obligations depend on the intended use, claims made, and operational context of deployment. The mappings below illustrate typical alignments for systems performing regulated decision support or compliance auditing, not automatic categorizations.

### EU AI Act (Regulation (EU) 2024/1689) [25]

When deployed to support or influence regulated determinations—such as compliance eligibility or safety-related recommendations—the system may fall within scope of Annex III as a potential High-Risk AI System. The exact classification depends on function, domain, and risk management plan.

### EU Medical Devices Regulation (MDR) 2017/745 [42]

If positioned as Software as a Medical Device (SaMD) for diagnostic or therapeutic support, the system could qualify as Class IIa under the MDR. Non-medical compliance or legal-advice deployments would not normally be within MDR scope.

### EU General Data Protection Regulation (GDPR) 2016/679 [43]

GDPR applicability is broad; the system design aligns with data-protection-by-design and accountability principles but does not replace a controller-level DPIA or lawful-basis analysis.

### United States (FDA) [44, 45]

Comparable deployments in clinical contexts could require a Class II 510(k) submission. In practice, AI/ML-based SaMD clearance typically takes 12–24 months and requires predetermined-change plans, validation, and QMS evidence. The specification here is intended only to align design discipline with those expectations, not to assert premarket eligibility.

### Regulatory note

These mappings are informative and non-binding. Classification outcomes may be confirmed through a qualified regulatory assessment based on the system's declared intended use, claims, and deployment environment. Implementers remain responsible for seeking appropriate CE marking, 510(k) clearance, or AI Act conformity assessment before production use.

### Implications for this specification

- Answer Receipts, provenance manifests, and audit trails provide the documentation and traceability artefacts suggested across frameworks.
- Human oversight and abstention rules (§7.2) meet AI Act and MDR oversight expectations.
- Privacy controls (§9, Appendix J.5) operationalize GDPR Arts. 5–25.
- Incident and revocation logging (§11, Appendix J.4) support post-market monitoring and accountability.
- Even outside formal scope, adopting these controls mitigates liability and supports audit readiness.

*

# LEVEL 2 — ARCHITECTURE (for architects)

## 2. Method & design principles

The method adopts verification-by-construction: only outputs proven to satisfy explicit contracts SHOULD be eligible to pass. Where schemas are crisp (identifiers, dates, jurisdictions, citation anchors), fail-closed gates SHOULD block emission unless every required check succeeds. Where semantics are looser, policy-governed routes quantify and bound residual risk rather than assuming it disappears. Inputs, prompts, retrieved text, and control signals are carried in separated channels to prevent prompt/route contamination, and every governed run is bound to a pinned snapshot of policies, models, and manifests (identified by SHA-256 content hashes) to preserve reproducibility. In practice this yields deterministic contracts at system boundaries, calibrated verifiers in gray zones, strict separation of data and instructions, and disciplined versioning so any decision can be replayed exactly as it was made. These principles align with established RAG foundations [11] and with modern dense-retrieval practice, including DPR-style indexing [12], [13].

## 3. Executable slots & gates (normative API)

An executable slot is a governed function with a stable interface: it accepts declared inputs, runs a sequence of policy checks, returns a governed emission decision, and emits the artifacts that justify that decision. Each slot exposes thresholds and a bounded risk budget so operators can tune sensitivity without changing code; every evaluation is time-boxed to avoid silent hangs—on timeout or ambiguity the slot ABSTAINS [9], [10]. Gates are concrete instantiations of these slots within a route: they enforce guarantees such as retrieval_quality (allow-listed shards within license, jurisdiction, TTL), grounded_answer (sufficient independent support, no contradictions, calibrated risk under FDR/FWER caps), training_promotion (no model update unless joint SLOs are met on a pinned snapshot), and privacy (block or redact when PII/PHI is detected). The API is normative: inputs are typed and versioned, checks are composable and side-effect free, decisions follow the return-state contract below, and artifacts include an evidence pack, verifier summaries, scope_diagnostics, and receipt hashes. Because slots are uniform, routes can be assembled, audited, and replayed deterministically: the same inputs under the same policy snapshot SHOULD yield the same decision and the same artifacts.

Return states (normative). ReturnState ∈ { PROMOTE_FULL, PROMOTE_LITE, ABSTAIN }.

- PROMOTE_FULL — Emit full Answer Receipt (fragment bodies permitted under policy).
- PROMOTE_LITE — Emit Lite Answer Receipt (selective disclosure; no fragment bodies; selectors may be omitted per policy).
- ABSTAIN — Do not emit; return rationale and scope_diagnostics.

Routing note. Lite vs full is determined by disclosure policy (disclosure_scope, fragment_mode) and proof/latency SLOs; degrading from FULL → LITE is permitted when proof size or timeouts trigger the degrade rule (§8, Appendix J.3).

Reason propagation (normative). Every decision SHOULD emit a machine-readable reason set. For ABSTAIN and PROMOTE_*, the canonical reason codes and human-readable summaries SHOULD be recorded in the receipt's scope_diagnostics and appended to escalation_trail (with timestamp, gate, and policy/version). Missing or unmapped reasons SHOULD be treated as errors.

### 3.1 Scope gate interface (expanded)

The scope gate is a deterministic function that inspects the candidate evidence set against the active policy and returns a pair—one boolean decision and a structured list of reasons. It rejects fragments that fall outside jurisdiction or effective-date windows, that violate license terms, that come from sources below the minimum trust tier, or that create temporal monoculture when diversity is required. It also fails fragments whose anchors are missing or broken, whose language does not match the route, or that are duplicates or near-duplicates; timeouts and malformed licenses are treated as explicit scope errors rather than soft warnings. On any scope failure the route abstains with reason "scope," and the receipt's scope_diagnostics section enumerates the offending categories with fragment counts, while PII/PHI and other sensitive fields are redacted. Because the interface is stable—scope_ok(fragments, policy) → (bool, reasons[])—operators can change policy without changing code, and auditors can replay exactly why an answer was allowed or blocked under a given snapshot.

## 4. The triptych (as mnemonics, not brands)

In this triptych, governance is treated as code (SHRDLU-like) [33], provenance as a supply chain (Memex-like) [29], and verification as a first-class surface (Xanadu-like) [32]. On the SHRDLU side, "slots" specify what the model may attempt and "gates" stipulate conditions that must hold before any emission. Each slot follows a disciplined lifecycle—inputs gathered, checks applied, decision taken, artifacts produced (logs, proofs, receipts)—while gates are fail-closed with auditable thresholds and explicit timeouts. Concretely, a scope/jurisdiction gate blocks out-of-policy material; a retrieval-quality gate filters shards that miss license, freshness, or trust criteria; a grounded-answer gate enforces support without contradiction; a privacy gate applies PII/PHI policy; and a training-promotion gate prevents tainted content from entering learning corpora. The result is generation constrained by policy up front rather than rationalized after the fact [11], [12], [13].

On the Memex side, provenance renders the evidence supply chain explicit: manifests curate allow-listed shards scoped by issuer, time window, jurisdiction, and license; each manifest is Merkle-anchored [21] so every cited fragment carries a version hash and inclusion proof for offline verification. To reduce monoculture and coordinated-source risk, citation sets are assessed with graph-independence $G_{\text{indep}}$ and domain-adaptive issuer caps.

On the Xanadu side, every material answer ships with a signed Evidence Pack (COSE/JOSE) [24] that discloses the Minimal Sufficient Evidence Set (MSES), counterfactual "what-would-flip" probes, verifier statistics, and the relevant proofs. Receipts may be issued in lite mode (hashes and selectors) or full mode (fragments per policy). Taken together, the triptych yields decision-level attestation: not a guarantee that sources are true, but cryptographic proof of what was checked and why the answer was allowed to ship.

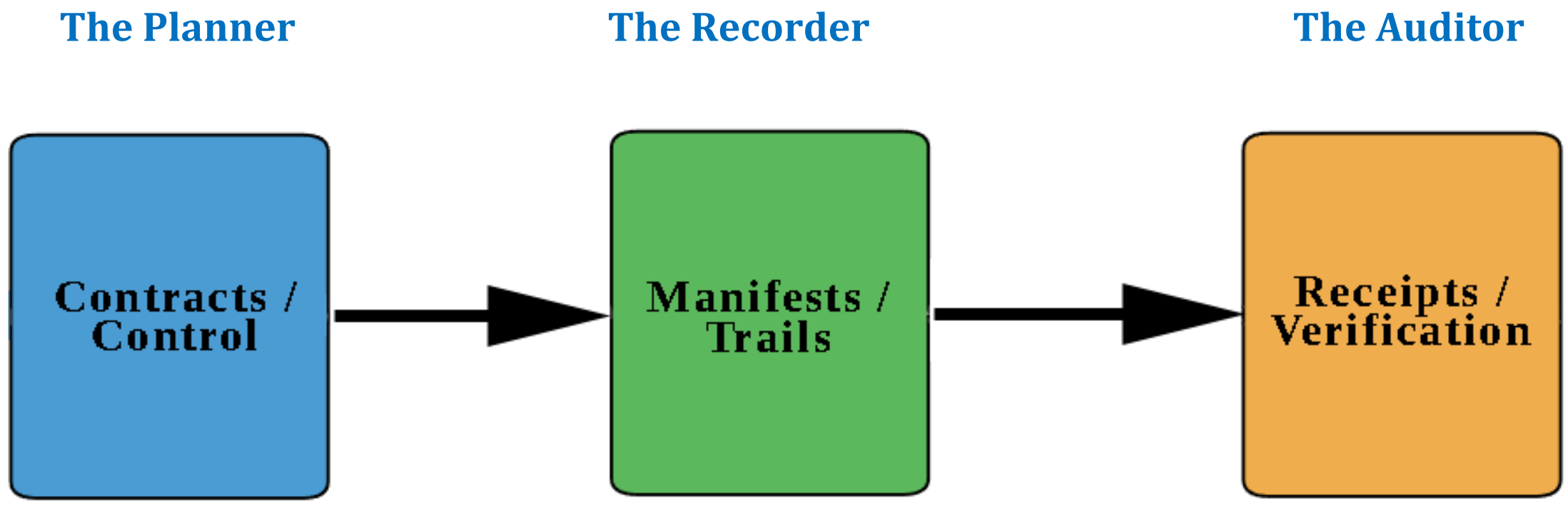

**Fig 1**. Triptych overview: Contracts → Manifests → Receipts

This architecture can be viewed as a three-layer stack: a Policy Layer (contracts, gates, NO-GO, independence) that governs the Generation Layer (retrieval, model, calibration, selective prediction), which in turn emits an Evidence Layer (manifests, Merkle trails, receipts). Policy flows top-down; evidence verification flows bottom-up for each decision.

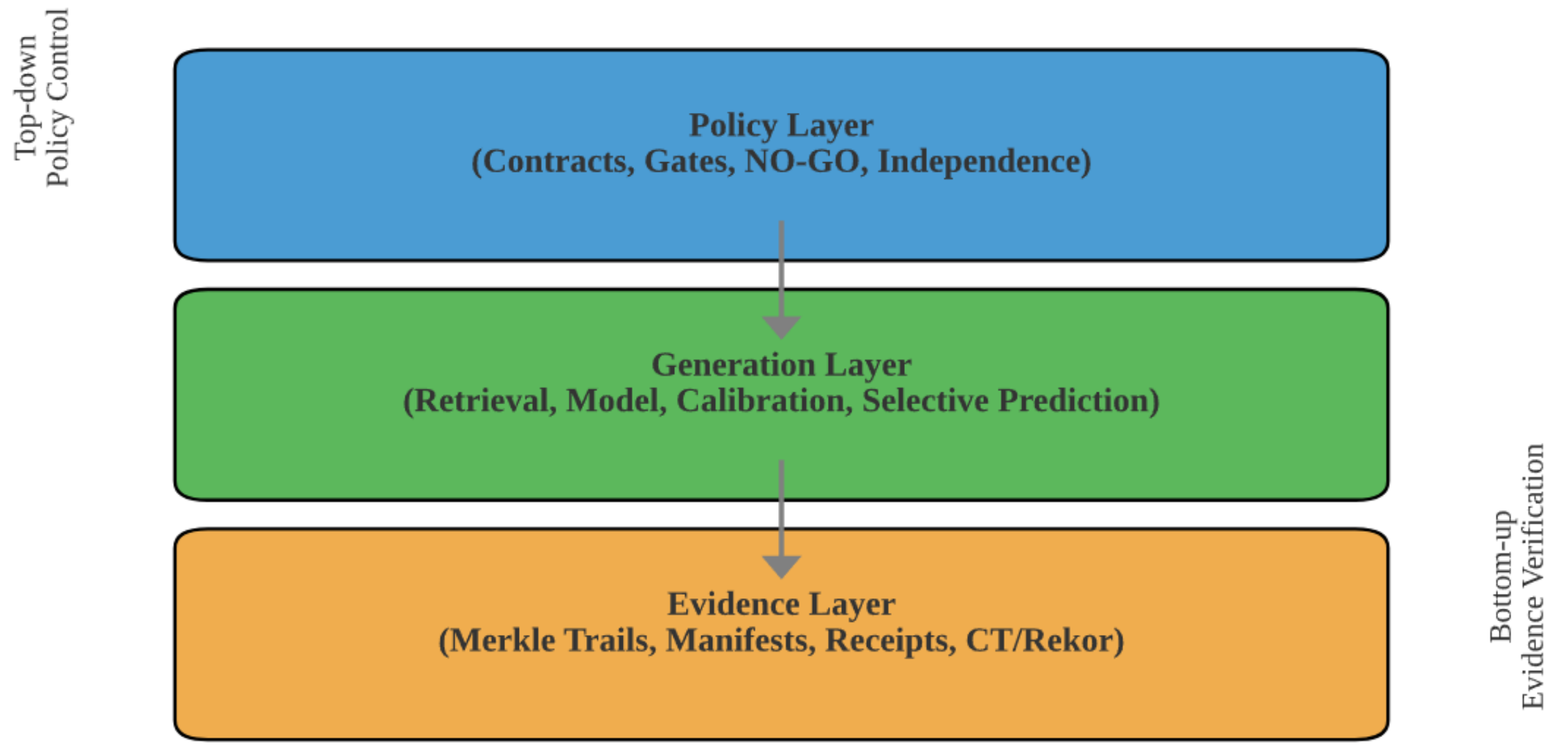

**Fig 2**. Architecture + Policy Stack: policy governs generation; generation emits verifiable evidence.

## 5. Threat envelope (prioritized, with mitigations)

**Scope.** Threats span both **data** and **control** planes. Prioritization uses a simple **Impact × Likelihood** rubric (**H/M/L**). Controls reference normative sections.

### 5.1 Prioritized threat list

1. Prompt injection (incl. RAG-mediated) — *Impact: H; Likelihood: H*
   Co-opts retrieved prose to issue hidden instructions, jailbreak policy, or exfiltrate secrets.
   Mitigations: strict data/instruction separation at router (§7); allow-listed tools only; selector-scoping + redaction (§9); scope gates (jurisdiction, TTL, license) (§7.1); cheap→small→heavy cascades (§10); pre-registered (optional) red-team on injection variants (§6, §D.6).
   Residual: advanced obfuscations → canaries (§7.4) and ABSTAIN on ambiguity (§7.2).
2. Data poisoning (incl. small-sample) — *Impact: H; Likelihood: M*
   Tainted shards, forged updates, or licit-but-biased texts; small-sample attacks succeed with near-constant poisons.
   Mitigations: issuer caps ≤ 50%, G_indep ≥ 0.70 (K=3) (§7.1–§7.1.2); temporal diversity on drift (§7.1); Merkle-rooted manifests + anchoring (§8); pinned snapshots + Promotion Receipts (§7.3, §9); adversarial canaries & shadow eval (§7.4).
   Residual: slow-roll bias → quarantine & rollback playbooks (§7.4).
3. Model extraction & membership inference — *Impact: H; Likelihood: M*
   Probes outputs/receipts to infer training data, manifest contents, or index boundaries.
   Mitigations: Lite Receipts by default; selector length cap; hashed fragments (§9); rate limits (per route/org) + query auditing (App. J.5); degrade on proof timeout (§8, J.3); disclosure scopes (internal-only/partner/public) (§9).

Residual: correlation across time → per-tenant budgets, anomaly detection, restricted disclosure scopes.

4. Temporal confusion (stale/mis-scoped dates/jurisdictions) — *Impact: M–H; Likelihood: M*
   Obsolete guidance appears current.
   Mitigations: effective-date/jurisdiction gates (§7.1); metadata-inference slot with ABSTAIN(needs_cataloging) when uncertain (§7.1.1); temporal-diversity rule on drift (§7.1).
   Residual: back-dated content → anchored manifests (§8).
5. Privacy leaks (PII/PHI) — *Impact: H; Likelihood: M*
   Fragments or receipts over-disclose.
   Mitigations: PII/PHI scan pre-finalization; mask or ABSTAIN (§9); selector-scoping policy; Lite Receipts default; anti-extraction limits (App. J.5).
   Residual: linkage risk → hash-only receipts / partner-only modes (§9).
6. Hallucinations (unsupported claims) — *Impact: M; Likelihood: M*
   Fluent but ungrounded content.
   Mitigations: ≥ 2 independent supports, contradiction screen (§7.2), FDR (BY) for supports & FWER (Holm) for contradictions (§10); abstention on weak margins (§7.2).
   Residual: corpus gaps → allow-list curation; needs_cataloging abstains (§7.1.1).
7. Supply-chain compromise (manifests/index/policy) — *Impact: H; Likelihood: L–M*
   Tampering with manifests, indices, or policy snapshots.
   Mitigations: SHA-256 content addressing (§7.3); Merkle-rooted manifests + anchoring (§8); HSM-backed signing; Promotion Receipts; two-person rule & hash-pinned promotions (§7.3); dual-channel revocation with fail-closed (§11, J.4).
   Residual: insider threat → role separation & immutable logs (App. B).
8. Byzantine signing / key compromise — *Impact: H; Likelihood: L*
   Compromised signer or inconsistent revocation state.
   Mitigations: KRN mirror + substrate revocation (dual check), REVOKED-PENDING-MIRROR fail-closed (§11, J.4); optional threshold ECDSA (App. I); audit of signing events; Lite fallback if proof SLOs breach (§8, J.3).
   Residual: mirror lag → freshness SLO + reconciliation (J.4).
9. Timing/side-channel on selective prediction — *Impact: M; Likelihood: L–M*
   Latency differences reveal thresholds or abstention logic.
   Mitigations: time-boxing, bucketed segment budgets (J.1); padding/jitter around τ-bands; constant-time paths for emit vs abstain where feasible; route-level p95 caps (§10, J.1).
   Residual: aggregate inference → restrict external timing telemetry; publish aggregates only.
10. Social engineering of human-in-the-loop (HITL) — *Impact: M–H; Likelihood: M*
    Adversaries influence reviewers to override gates.
    Mitigations: dual control for high-stakes releases (§7.2); structured checklists & escalation_trail logging (§3, §9); role separation + cool-down windows (App. B).
    Residual: fatigue/pressure → rate caps, rotation, post-hoc audits of overrides.

### 5.2 Operational stance

All retrieved text is untrusted by default. Routers strictly separate data from instructions; policy gates enforce scoped provenance, contradiction screening, independence/triangulation, and privacy before emission (§7). Residual risk is non-zero;

the aim is to bound blast radius, surface verifiable evidence, and enable rapid quarantine and rollback—not to promise invulnerability.

Implementation notes (non-normative):

injection hygiene (schema-constrained outputs; deny unsafe tool calls), poisoning hygiene (anchored manifests; shard-update review), extraction hygiene (Lite receipts; per-tenant budgets), timing hygiene (pad near τ; limit per-check timing export).

## 6. Red-team protocol (pre-pilot)

Purpose. Prior to any pilot, an independently led, pre-registered (optional) red-team exercise SHOULD stress the system to quantify defensive efficacy under adversarial pressure.

Scope & snapshot discipline. Evaluation is conducted on a frozen, hash-pinned snapshot of all governed components: models (either self-hosted *weights hash* + config + container/image digest, or API model with provider, model ID, version string, decoding parameters and RNG seed, plus evaluation date window), prompts/routers, retrieval indices, verifiers, policies, thresholds, and provenance manifests. Identifiers (e.g., SHA-256) for each artifact SHOULD be recorded in the preregistration package (see §G.2).

Attack families. The exercise SHOULD cover at least the following classes:

1. Prompt injection, including RAG-mediated variants [19].
2. Citation stuffing / retrieval padding.
3. Paraphrase laundering (semantic restatement to evade checks).
4. Threshold probing (boundary gaming of emission/abstention).
5. Source collusion (coordinated or monoculture supports).
6. Temporal confusion (mis-scoped effective dates/jurisdictions).
7. Model extraction & membership inference (probing outputs/receipts to infer training or manifest contents).

Attack sets SHOULD be power-planned per class (typically 800–1,200 attempts) so outcome estimates have meaningful confidence intervals. Generation of attacks SHOULD mix internal adversaries, scripted agents, and domain experts; execution SHOULD be against the frozen snapshot and against strong baselines (B1, B2 as defined in §D.7).

Baselines (reference). B1: RAG + calibration + human-handoff; B2: RAG + guardrails (no receipts). Both tuned on the same frozen snapshot (§D.7).

Primary metrics.

- ASR (attack success rate).
- Q2S (queries-to-success; median).
- TTD (time-to-detect; p95 minutes) under live monitoring.
- Benign p95 latency (interactive routes).

Passing criteria (vs strongest baseline, 95% confidence).

- ASR ↓ ≥ 30%,
- Median Q2S ↑ ≥ 3×,
- p95 TTD ≤ 15 min,
- Benign p95 latency ≤ 900 ms (defenses SHOULD NOT "pass" by slowing everything down).

Design controls.

- Sequential testing MAY be used with pre-declared α-spending (O'Brien–Fleming) [4] to stop weak classes early or intensify promising ones.
- Randomization & power: class-level sample sizes and interim looks SHOULD be pre-registered (optional) (see §G.2; power guidance in §D.10).
- No leakage: tuning data SHOULD be disjoint from held-out evaluation; snapshot changes during evaluation are forbidden. Any deviation SHOULD be logged as an amendment in the preregistration.

Logging & receipts. Every attack, detection, and mitigation SHOULD be receipt-logged (input/query hash; route/policy versions; evidence/proofs; verifier outcomes; reasons; timestamps). Logs SHOULD enable offline replay and adjudication.

Operational safeguards. Live-ops telemetry (guard-railed) SHOULD measure detection and trigger quarantine/rollback per incident policy. Rate limits, disclosure modes (Lite Receipts, selector scoping), and redaction SHOULD be enforced during testing to bound extraction/membership risks (see §9, §10).

Outcome. The intended outcome is not invulnerability, but a quantified reduction in attacker efficiency and tighter detection windows demonstrated against the baselines under controlled, reproducible conditions.

## 7. Operational conditions (must hold)

This section defines the non-negotiables for safe operation: what evidence may be retrieved, how answers must be grounded, when updates can be promoted, and how drift is detected and contained. Treat this as the runtime contract that keeps policy, performance, and provenance aligned.

### 7.1 Retrieval Requirements (allow-lists, caps, G_indep)

All retrieval SHOULD be confined to allow-listed shards whose entries carry, at minimum: license, TTL, trust tier, jurisdiction, and effective-date metadata.

Independence controls (applied at support time):

- Issuer cap (default 50%) — No single issuer may account for more than 50% of the supporting fragments for a given answer (i.e., for every issuer *i*, share_i ≤ 0.50).

- Graph independence — The support set SHOULD satisfy G_indep ≥ 0.70 with default hop radius K = 3 (i.e., no excessive shared upstream sources within three hops in the provenance graph).

Temporal diversity (applies when a topic is flagged as drifting): supporting evidence SHOULD span ≥ 2 time windows, each ≥ 6 months in duration, with midpoints separated by ≥ 12 months (ensuring temporal separation). These controls are enforced in addition to the scope gate (jurisdiction, license, TTL).

Small-sample poisoning gate (normative). Before emission, supports SHOULD satisfy: (i) issuer diversity ≥ 3 in the candidate pool; (ii) MSES monoculture cap: in the final MSES, ≤1 fragment per issuer *(Phase B+)*; (iii) G_indep ≥ 0.70 (K = 3) and issuer share ≤ 50%; (iv) on routes flagged "drifting," enforce temporal diversity (≥ 2 windows, each ≥ 6 months; midpoints ≥ 12 months). Failing any sub-gate → ABSTAIN('insufficient_diversity') with scope diagnostics in the receipt.

Worked pass/fail (surfaced early).

- Fail: $f_1$ (EUR-Lex A, 2020), $f_2$ (MDCG B, 2020), $f_3$ (White Paper C citing B, 2020).
- Pairs = 3; shared upstream within K = 3 → shared = 1, so G_indep = 1 − 1/3 = 0.667 (fail).
- Issuer cap: two of three supports share one issuer → 2/3 = 66.7% > 50% (fail).
- Temporal diversity: all supports lie in the same window (fail).
- Pass: Replace $f_3$ with an Official Journal fragment from 2022 that does not cite B (distinct issuer).
- Now shared = 0 → G_indep = 1.0 (pass); max issuer share = 1/3 ≤ 50% (pass); two disjoint windows (2020 vs 2022) with midpoint separation ≥ 12 months (pass).

These requirements align with established retrieval-augmented generation practice [11].

#### *7.1.1 Metadata-inference slot (for messy corpora)*

If jurisdiction or effective dates are missing or ambiguous, a metadata-inference slot MAY estimate them with calibrated confidence. When confidence falls below threshold, the route SHOULD ABSTAIN with reason needs_cataloging and return pointers to the offending documents for ops triage. Inferred metadata SHOULD NOT override explicit authoritative tags; they enable safe abstention rather than silent acceptance.

#### *7.1.2 G_indep algorithm & example*

Given a support set S = $\{f_1,\ldots,f_n\}$ with issuer and upstream-citation lists and K = 3 by default:

1. Build the provenance graph.
2. Flag any pair that shares a common ancestor within K hops.
3. With pairs = n(n−1)/2 and shared the count of flagged pairs, compute
4. G_indep = 1 − shared / pairs and gate at G_indep ≥ 0.70.

Receipts SHOULD report K, the flagged pairs, and the independence decision. Any use of K < 3 SHOULD be explicitly justified by policy and recorded in the receipt's policy/version fields.

### 7.2 Grounding Rules (supports, contradictions, thresholds)

An answer MAY be emitted only when all of the following hold:

- Support: At least two independent supporting fragments pass verification and independence checks.
- Contradictions: No detected contradictions to the asserted claim(s).
- Justification threshold: The composite justification score meets or exceeds the configured threshold under multiplicity control (e.g., FDR for supports; FWER for contradictions).
- Scope & licensing: All fragments are in-policy (jurisdiction, effective dates, license/TTL, trust tier).

Verification SHOULD proceed in a cascade from cheap → small → heavy checks. When margin signals are weak or verifiers disagree within the configured band, the route SHOULD ABSTAIN. For designated high-stakes queries, human spot-review SHOULD be required before release [38, 39].

### 7.3 Promotion Discipline (pinned snapshots, SLOs)

Model or index updates SHOULD NOT be promoted on the fly. Promotion SHOULD occur only from a pinned evaluation snapshot that meets joint SLOs (coverage, contradiction control, latency, provenance completeness). Each promotion SHOULD emit a signed Promotion Receipt capturing the snapshot identifiers (e.g., SHA-256 content hashes), policies, thresholds, and verification outcomes. Snapshots, decoding parameters, and seeds SHOULD be recorded to ensure exact replay.

#### *7.3.1 Supply-chain & signing hardening (normative)*

Signed CI + SBOM. All manifest artefacts SHOULD be produced via a signed CI pipeline; deployables SHOULD be code-signed and accompanied by an SBOM. Artefacts and SBOMs SHOULD be retained for D1/D2 ops audits.

Quorum promotion. Every manifest promotion SHOULD emit a Promotion Receipt signed by a quorum (t-of-n or documented multi-sig). The receipt SHOULD include approver KIDs and a Promotion SHA-256 over the promoted set.

Two-person rule. Production promotions SHOULD require two-person approval for shards above the configured impact threshold (domain-specific).

Dual-channel revocation. Verification SHOULD check both the local KRN and any substrate KRN; on mismatch or freshness breach, fail-closed and emit an incident receipt (see §9, §11, App. I/J).

### 7.4 Drift Monitoring (canaries, quarantine)

After promotion, rolling canaries and shadow evaluations SHOULD run on fresh traffic to detect drift or transfer attacks. If operational gates begin to fail, affected shards or routes

SHOULD be quarantined or automatically demoted pending investigation. Incident records SHOULD include detection time, blast radius, and mean time to remediation to support operational SLOs. Re-promotion REQUIRES restored metrics under a fresh snapshot and a brief A/B sanity check.

## 8. Provenance manifests

Provenance manifests are maintained at shard scope—each shard defined by *(issuer, corpus, jurisdiction, effective-date window)*—and list per-document fields *(doc_id, version_hash, license, anchors, trust_tier)*. The ordered listing is committed to a Merkle root [21] and anchored to a public timestamp so auditors can later verify inclusion for any cited fragment. To keep proofs compact and verification fast, implementers cap shard size (e.g., ≤ 64k leaves), use sparse Merkle trees (SMTs)—keyed by a document-ID hash over a fixed keyspace with default empty leaves—to support both inclusion *and non-inclusion* proofs with aggregated paths, and archive superseded roots; manifests record both issuer and author. Multiproof aggregation exploits shared path segments so that a bundle of fragments verifies with near-logarithmic overhead. Caps ordinarily apply at the issuer level, with optional author-level caps to mitigate "ghost-network" risks.

Normative proof SLO & degrade rule. Proof size and responsiveness are enforceable:

- SLO (report in receipts): proof_size_p90 < 64 KB, proof_size_p99 < 256 KB, and proof_verify_p95 ≤ 200 ms on the governed route's reference hardware.
- Per-request guardrails: if a proof exceeds the bound or times out (implementation default: proof_timeout_ms = 300), the route SHOULD either emit a Lite Receipt (if other supports satisfy policy) or ABSTAIN('proof_timeout').
- Receipts SHOULD carry proof_size_bytes, proof_timed_out, and the proof policy version used to adjudicate the degrade.

Anchoring & signing substrates (non-normative guidance). This specification is substrate-agnostic: receipts SHOULD be portable COSE/JOSE objects with Merkle multiproofs [21] verifiable offline by a standalone verifier; substrate choice SHOULD NOT alter receipt format or verifier logic. Operators MAY anchor roots via (a) on-prem HSM-signed sidecar logs with periodic notarization, (b) CT/Sigstore/Rekor-style append-only logs [22], or (c) certified-data protocols with threshold ECDSA (e.g., ICP/IC-Guard). See Appendix I for non-normative options and trade-offs. For SMT background and proof composition, see Merkle Tree Proofs (RFC 9162) [36].

## 9. Signed Answer Receipts

Purpose. A Signed Answer Receipt records what was checked and why emission was allowed, without asserting source truth. Receipts are portable, offline-verifiable evidence objects.

### 9.1 Receipt contents (normative)

Each receipt SHOULD include at minimum:

- Answer & policy context
  - answer_hash
  - route_version, contract_version
  - disclosure_scope ∈ {internal-only, partner, public}
  - fragment_mode ∈ {hash, digest, full}
- Evidence & verification
  - MSES (Minimal Sufficient Evidence Set) with counterfactual flips and per-fragment reasons
  - verifier_stats (e.g., Holm/BY adjusted p-values; measured m_eff when available)
  - scope_diagnostics (license, TTL, jurisdiction, effective-date checks)
  - abstain_reason (if applicable)
- Independence & diversity diagnostics
  - mses_issuer_counts
  - g_indep_value (default K=3)
  - temporal_windows (if topic flagged drifting)
  - poisoning_gate_pass ∈ {true,false} and poisoning_gate_reason when false
- Proofs & signing
  - Merkle multiproofs for all cited fragments
  - KID (signer id), alg
  - Promotion & signing metadata: signing_scheme (single | multi-sig | threshold with n,t), promotion_digest (SHA-256 over promoted manifest set), and promotion_approvals (ordered approver KIDs + timestamps)

- Timing-hardening summary

  - proof_latency_histogram (aggregate bins only, e.g., p50/p95/p99)

  - jitter_policy (descriptor of padding/windowing in effect)

### 9.2 Selective disclosure & privacy (normative)

Receipts use selective disclosure by default: cited fragments are represented as hashes/digests + selectors. A PII/PHI scan runs pre-finalization; if sensitive data is detected, the system SHOULD mask it or ABSTAIN with a privacy reason. Selector-scoping SHOULD prevent rare-token or uniquely identifying n-gram leakage. Linkability risks SHOULD be noted in the receipt. Internal-only mode keeps receipts verifiable on-prem while disabling external sharing.

### 9.3 Portability (normative)

Receipts SHOULD be portable COSE/JOSE objects that a standalone, offline verifier can validate without contacting any external log. Substrate choice SHOULD NOT alter receipt fields or verification logic. When threshold/multi-signature is used, scheme parameters SHOULD be encoded in signing_scheme, and approvals included in promotion_approvals SHOULD suffice for offline verification.

### 9.4 MSES computation note

For Phase A, MSES is computed via greedy backward elimination with up to three rechecks to stabilize the minimal set; receipts report Δ-scores and a fragility index. (Later phases may substitute a heavier verifier so long as schemas/SLOs hold.)

Illustrative example: Fig. 3 shows initial evidence vs. stabilized MSES; removed nodes (dependence or adjusted-threshold failures) are grey “X”, retained supports green, contradictions (Holm-controlled) [1] pale red.

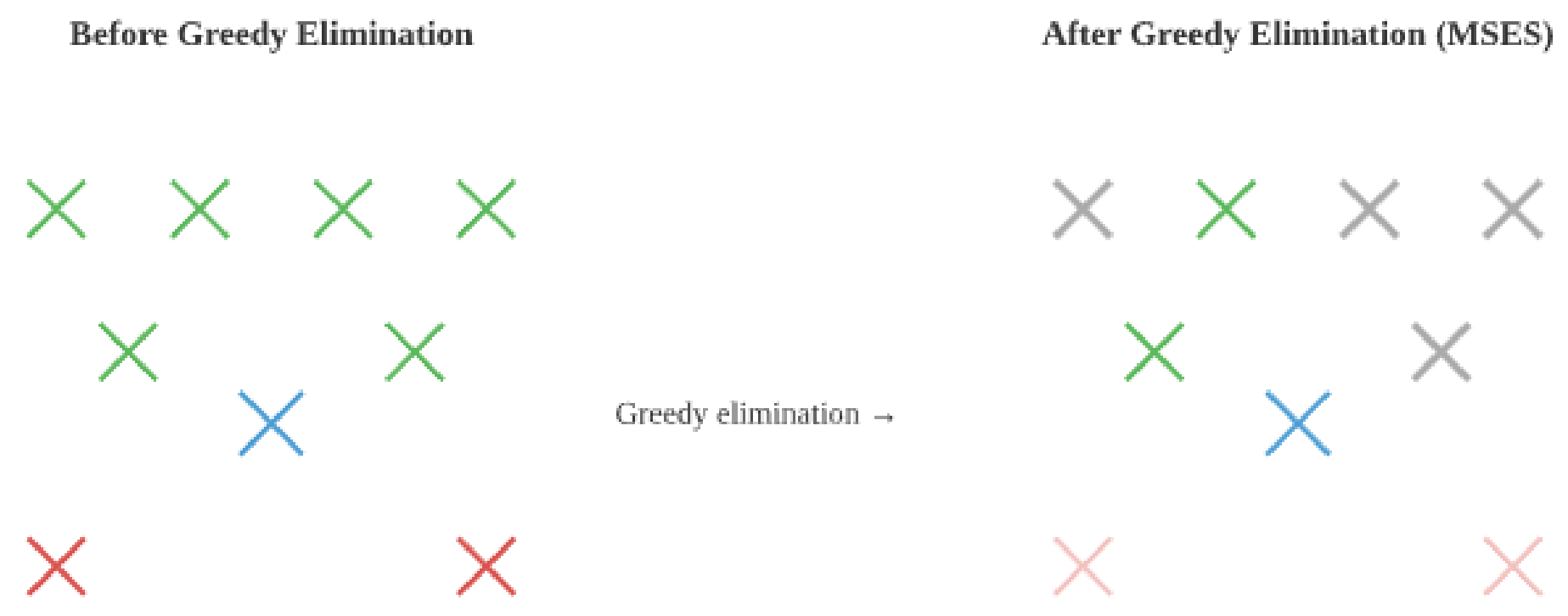

**Fig. 3**. MSES extraction: minimal evidence set after greedy removal under independence and Holm-adjusted multiplicity

### 9.5 Side-channel hardening (normative)

Verification and proof generation operate within bounded windows with optional jitter. Per-ticket raw timings SHOULD NOT be exposed; only aggregate proof_latency_histogram is included. On proof timeout the system SHOULD degrade to Lite Receipt or ABSTAIN('proof_timeout') and record the outcome.

### 9.6 Lite Receipts

Lite Receipts omit fragment bodies (and may omit selectors), shipping only scope diagnostics, inclusion multiproofs, route/contract versions, timing summary, and a signature—substantially smaller but still offline-verifiable.

### 9.7 Revocation checks (normative)

Verifiers SHOULD check the signer KID against the local KRN and, when applicable, the substrate’s revocation state. Verifiers SHOULD perform dual-channel checks; on mismatch or mirror SLO breach they fail-closed, emit an incident receipt referencing promotion_digest, and mark the original REVOKED-PENDING-MIRROR. See §11.2–§11.3 and Appendix I for mirror SLOs and anchoring options.

## 10. Uncertainty-aware evaluation & budgets

The verifier’s core job is to distinguish good claims from bad ones. However, no verifier is perfect. This section describes the statistical controls used to manage verification errors so they do not compromise output quality.

Verifier engine rationale. A DeBERTa-class entailment model [14] is used as the verification backbone because it (i) achieves competitive accuracy on standard NLI benchmarks (e.g., MNLI, RTE) as reported in [14]; (ii) its disentangled attention architecture provides precise signals for support and contradiction checks; and (iii) simple post-hoc temperature scaling yields reliable calibration (i.e., stated confidence ≈ empirical accuracy) in pilot testing [8]. To avoid dependence on a single model's idiosyncrasies, any proposed verifier swap (e.g., to a RoBERTa-class model) SHOULD meet the same calibration and accuracy targets under the selective-prediction protocol [9, 10].

Controlling system-level risk. The core risk metric is confident-error@$\tau$: the percentage of emitted answers that are wrong despite the system claiming confidence $\geq \tau$. Because one answer can involve dozens of correlated checks, naïvely treating $p$-values as independent would understate risk. Explicit multiplicity controls are applied:

- Contradictions → FWER control (Holm) [1]. Even one false contradiction invalidates the output; Holm's step-down procedure controls the family-wise error rate.
- Supports → FDR control (Benjamini–Yekutieli) [2]. Supporting fragments are often dependent; B-Y controls FDR under arbitrary dependence. *(Storey $q$-values [3] are recorded only as exploratory signals.)*

Dependence among checks is summarized by the effective number of tests $m_{\text{eff}}$; as dependence rises, risk budgets are tightened to maintain the desired margin.

Auditable decisions. These controls are coupled to behavior:

- If calculated error bounds are exceeded, the route abstains or tightens criteria.
- If calibration drifts, the internal threshold $\tau$ is adjusted within a safe range.

All verification decisions log the active statistical budgets and the adjusted $p$-values so receipts show what passed and under which uncertainty regime [9, 10].

Latency budget (per route). Each governed route SHOULD declare a latency-budget table allocating p95 wall-clock across: retrieval; cheap/small/heavy verifiers; proof assembly; and signing. The default end-to-end target is p95 ≤ 900 ms. Defaults for timeouts and retries are centralized (Appendix J — Operational defaults & SLO tables); routes MAY tighten but SHOULD NOT exceed these defaults without logging an SLO change in the route manifest.

### 10.1 Heavy verifier enabling (adaptive policy & budget)

Triggers (pre-registered) (optional). Heavy verification MAY be invoked only when at least one holds: (i) verifier disagreement; (ii) confidence margin within band [0.55, 0.75]; (iii) high-stakes flag on the claim; (iv) adversarial indicators (ICL score $\geq \tau_{ICL}$ or retrieval anomaly); (v) deterministic random-sampling for monitoring (per-route default 10 %, seeded and logged). Triggers SHOULD be deterministic and logged.

Per-request limiter. At most one heavy-verifier call per request (≤ 1); additional heavy calls in the same request are forbidden and blocked by a circuit-breaker.

Route-level cap. For each governed route, the rolling 7-day sliding window share of requests invoking heavy verification SHOULD be ≤ 15 %. When this cap would be exceeded, the system SHOULD atomically reserve/verify capacity and either:

• Degrade to Lite Receipt if remaining supports suffice (proofs + policy checks pass); or
• ABSTAIN("budget_exhausted") with reasons if Lite Receipt is not permissible.

Incident mode. During an attested incident (drift/anomaly declared by Operations; ticketed; time-boxed), routes MAY exceed the 15 % cap (up to 100 % in attested-async). Incident start/stop, scope, and duration SHOULD be logged; reversion to standard caps is mandatory at closure, and a post-mortem is included in § G.5 / § G.11.

Conditional-power tie-in. If low_conditional_power() is True (per § G.4), the random-sampling trigger rate MAY be increased up to, but not beyond, the route cap. Incident mode is not implied by low power.

Latency constraints & retries. Interactive p95 ≤ 900 ms SHOULD be maintained. Stage timeouts and single retries follow Appendix J (defaults: 300 ms per stage, max 1 retry, idempotent). A circuit-breaker SHOULD prevent repeated heavy invocation within the same request. Each route SHOULD publish a latency-budget table (ms slices for retrieval, verifiers [cheap/small/heavy], proof assembly, signing) and demonstrate that typical heavy-enabled paths fit within the table.

Accounting & audit. Every heavy decision SHOULD log: request_id, route_id, user/session class, active trigger set, whether budget permitted it (with reservation id), elapsed stage times, and impact on BY/Holm risk budgets. Budget counters are monotone within the 7-day window (true sliding window by timestamp). Reasons returned by ABSTAIN / PROMOTE_* SHOULD appear in receipts' scope_diagnostics and escalation_trail.

Substrate independence. This policy is independent of the anchoring substrate (see Appendix I). On degrade, receipt mode is Lite Receipt (scope + inclusion proofs + signature).

Decision **pseudocode** (limiter + cap + degrade + latency budget)

```
def decide(claim, fragments, policy, route_id, now, budget_ms):
    # Scope & retrieval
    ok, reasons = scope_ok(fragments, policy)
    if not ok:
        return ABSTAIN("scope", reasons)

    # Stage 1: cheap checks (budgeted)
    s1 = [cheap_nli(f, claim, timeout=policy.s1_timeout_ms)
          for f in fragments if not is_timeout(f)]
    cand = select_topk(s1, k=policy.K1)
```

```
    # Stage 2: small NLI (budgeted)
    s2 = [small_nli(f, claim, timeout=policy.s2_timeout_ms) for f in cand]
    supp = [f for f, p in s2 if p.support and not p.contradict]
    if len(supp) < 2:
        return ABSTAIN("insufficient_support")

    # Multiple-testing budgets
    if not fdr_ok_BY(supp, q=policy.q) or not fwer_ok_Holm(cand, alpha=policy.alpha):
        return ABSTAIN("risk_bounds")

    # Heavy enabling with limiter & cap (atomic capacity check)
    need_heavy = heavy_triggered(claim, fragments, policy) or low_conditional_power()
    heavy_allowed_req = (per_request_heavy_calls() == 0)  # ≤1 per request
    share = route_heavy_share(route_id, window_days=7, now=now)
    incident = incident_mode(route_id)
    have_capacity = (share < 0.15) and reserve_heavy_slot(route_id, now)  # atomic
    do_heavy = need_heavy and heavy_allowed_req and (have_capacity or incident)

    if do_heavy and not fits_latency_budget(budget_ms, policy.heavy_cost_ms):
        do_heavy = False  # fall through to degrade/abstain logic

    if do_heavy:
        mark_heavy_used()  # increments per-request counter
        hv = [heavy_nli(f, claim, timeout=policy.hv_timeout_ms) for f in cand]
        if any(r.contradict for r in hv):
            return ABSTAIN("heavy_veto")

    # Independence & issuer caps
    if g_indep(supp, K=policy.K) < 0.70 or issuer_cap_exceeded(supp, cap=policy.cap):
        return ABSTAIN("independence_or_cap")

    # Proof policy with degrade to Lite Receipt when proofs slow/oversize
    if merkle_proofs_missing_or_slow(supp, timeout_ms=policy.proof_timeout_ms):
        if lite_receipt_ok(supp):
            return PROMOTE_LITE(supp, reason="proof_timeout_or_size")
        return ABSTAIN("proof_timeout_or_size")

    # If heavy was needed but capacity exhausted (no incident), degrade or abstain
    if need_heavy and not incident and not have_capacity:
        if lite_receipt_ok(supp):
            return PROMOTE_LITE(supp, reason="heavy_budget_exhausted")
        return ABSTAIN("budget_exhausted")

    return PROMOTE_FULL(supp)
```

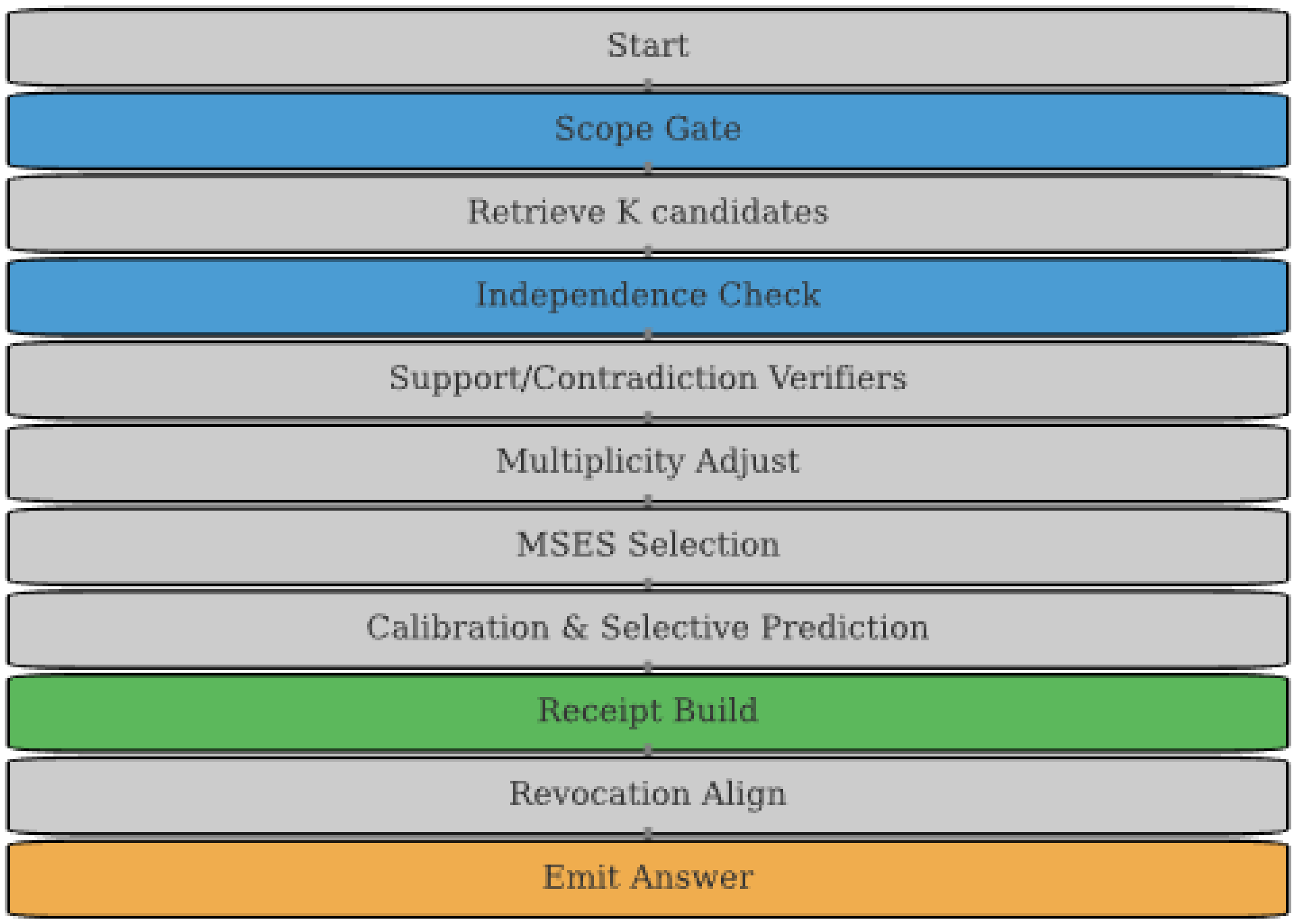

**Fig 4**. Decision flow: gates, multiplicity, calibrations, receipts.

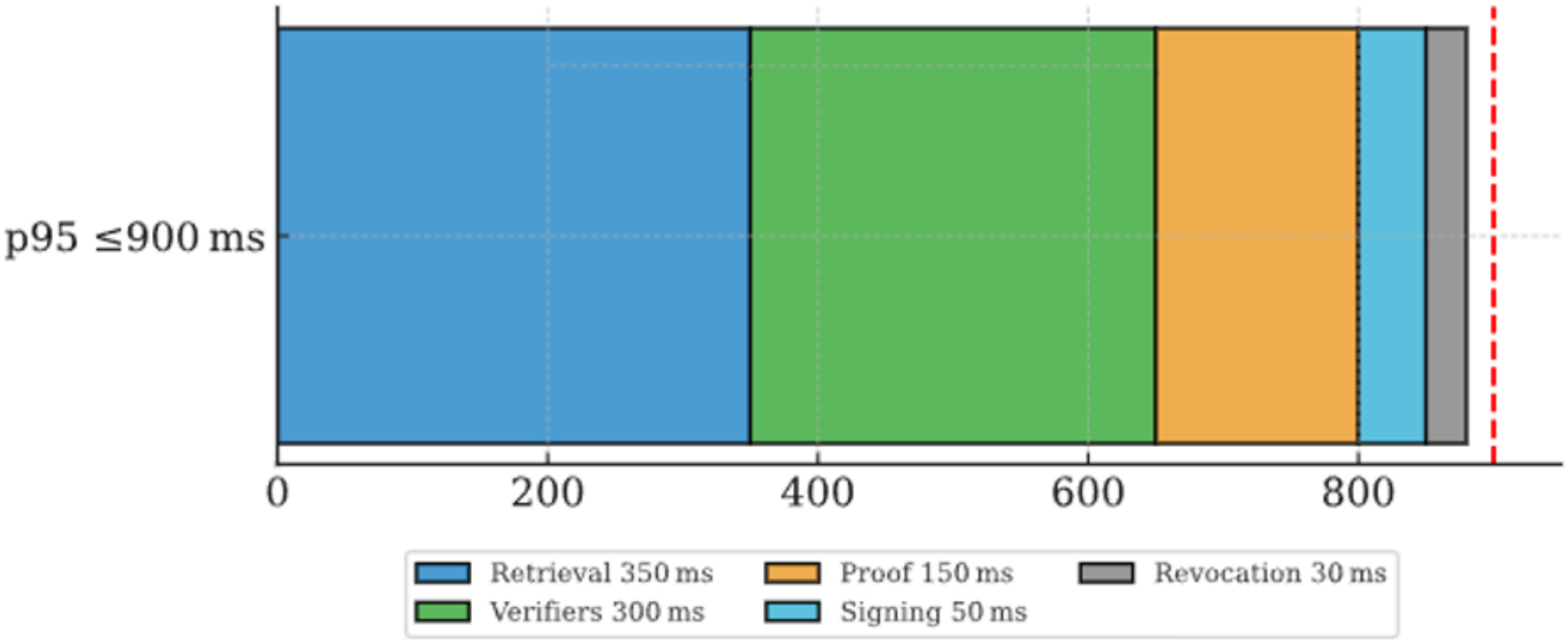

**Fig 5**. Latency Budget Breakdown by component

Implementation notes

- heavy_triggered(...) implements the trigger set above; keep it deterministic and logged.
- per_request_heavy_calls() SHOULD be zero at entry and incremented once at first heavy call.
- route_heavy_share(...) returns the 7-day sliding ratio (heavy_requests / total_requests) for the route.
- fits_latency_budget(...) checks the route's declared budget table (retrieval / verifiers [cheap/small/heavy] / proofs / signing) and remaining headroom vs. policy.heavy_cost_ms (see Appendix J).

- Incident state is set/cleared by operations with receipts and timestamps (see §G.6, §G.11).
- All returned reasons are surfaced in receipts' scope_diagnostics and escalation_trail.

Deliverables (pilot). The D1 Ops Report (Appendix G.5) SHOULD include (i) the per-route latency budget table, (ii) observed p50/p90/p95 by stage, and (iii) heavy-share over the 7-day window with any incident intervals highlighted, demonstrating conformance with this policy.

## 11. Governance-layer threat model

Key compromise: HSM, rotation ≤12 months, threshold signatures; role-separated keys (manifests vs receipts). Insider risk: split authority (Compliance + Security) and external monitors.

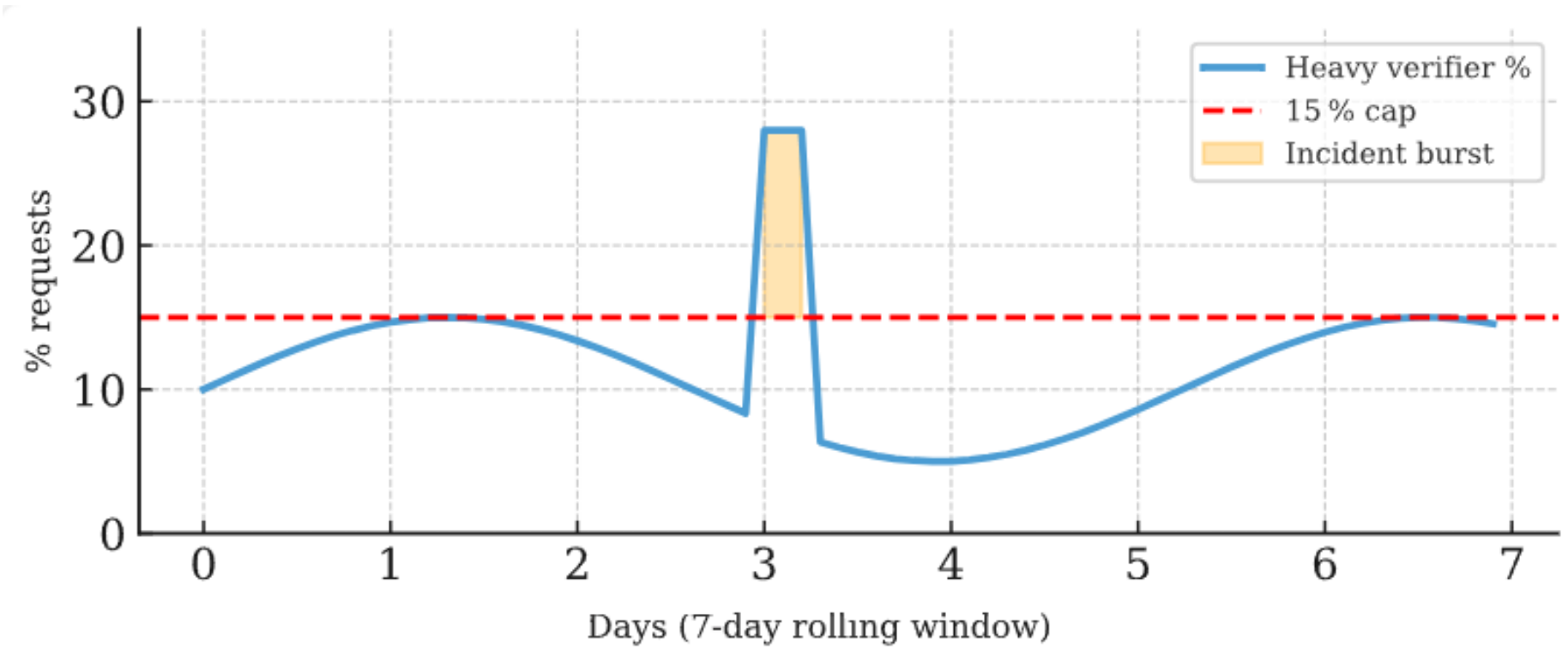

**Fig 6.** Heavy verifier governance: 15% cap & burst

Log integrity (Phase C): CT-style append-only with cross-publishing. Verifier poisoning: reproducible builds; canary suites; model provenance.

### 11.1 Internal-only attestation mode

In internal-only attestation mode, every receipt is signed by organization-owned keys in an on-prem HSM; no external transparency log is used. Receipts remain fully verifiable by the internal offline verifier (COSE/JOSE + Merkle multiproofs), while external parties receive decision summaries rather than raw evidence. Disclosure is controlled by policy (hash/digest selectors by default), with PII/PHI redaction enforced before any sharing. Key rotation and revocation follow the KRN runbook, and all signing events are immutably logged for internal audit. This mode minimizes legal/operational exposure while preserving decision-level accountability. If public verifiability is later required, migration to a transparency substrate occurs in Phase C without changing receipt format.

### 11.2 Receipt invalidation & key compromise handling

Upon suspected or confirmed key compromise, a Key Revocation Notice (KRN) covering the window $[t_0, t_1]$is issued and promotions are immediately frozen. Affected keys are rotated, and all receipts signed by the compromised key (scoped by key ID and time window) are flagged UNDER REVIEW. Where underlying evidence remains intact, receipts are re-attested and re-signed; otherwise they are marked UNREISSUABLE with a documented rationale. An incident bulletin discloses scope, counts, and MTTR. Verifiers must reject any receipt bearing a revoked key ID unless a valid re-signature is present. Each receipt includes key ID, algorithm, and certificate fingerprint to enable deterministic validation against the current KRN.

### 11.3 Revocation alignment (substrates)

To avoid split-brain trust, any revocation or key rotation issued by the chosen substrate (e.g., CT/Rekor entry, ICP canister KRN) is mirrored into the proposal's canonical KRN stream. While CT/Rekor provide append-only transparency [22], receipt verifiers SHOULD consult both channels—the substrate's revocation state and the local KRN—accepting a receipt only if it passes both checks. Mirroring is near-real-time and auditable: each KRN entry records source, hash, timestamp, and linkage to the substrate event. On mismatch or fetch failure, verifiers fail-closed (treat as revoked) and emit a diagnostic. Receipts include substrate identifiers (log ID/canister ID) and the signer's key ID to bind checks deterministically. Periodic reconciliation jobs surface drift; incidents trigger backfill and re-attestation where feasible.

*

# DISCUSSION

## D.1 Scope & claims revisited

The program attests process, not truth: receipts document what was checked and under which thresholds, without asserting that sources are correct or that model internals are "understood." Usefulness is evaluated alongside accuracy through a helpful-abstain measure (practitioner Likert ratings, median ≥4/5 with confidence intervals), so judgment rests on safe task support rather than omniscience.

## D.2 Policy-checked vs policy-enforced

Policy-Checked. This is a probabilistic or complex rule applied to open-ended tasks (like summarization or Q&A). The system evaluates the policy (e.g., "Is this support strong enough?"), but if the signal is weak, it is allowed to abstain (refuse to answer) instead of guessing. The failure is not a hard block but a transparent admission of uncertainty. Risk Profile: Managed and bounded risk. The residual risk is "explicit and auditable.

Policy-Enforced. This is a binary, deterministic rule. The output either passes the rule perfectly (e.g., "Is the output valid JSON format?") or it is instantly blocked (fail-closed). This is possible only in constrained environments (schemas, forms). Risk Profile: Low risk, high certainty.

### D.3 Error propagation under dependence

Compounding from pair→claim→answer is analyzed under correlated verifier errors, with Monte-Carlo overlays quantifying risk using copulas and beta–binomial overdispersion [6], [7]. Agreement statistics (κ) and effective test counts (m_eff) are bootstrap-estimated with 95% confidence intervals and reported per snapshot, and alternative decision rules are compared under a pre-registered plan (optional).

### D.4 Anti-gaming metrics

Justification quality combines relevance, MSES-based minimality, triangulation, and graph independence (G_indep). Antagonistic tactics—padding, paraphrase laundering, source collusion—are countered by marginal MSES tests and two concrete controls: threshold jittering (±0.02 per session to reduce boundary gaming) and canary tests (1–2% of claims, refreshed monthly) to surface emerging exploits. Goodhart effects are acknowledged; metrics are monitored for drift and retuned only under pre-declared procedures.

### D.5 Computational feasibility

Routes are engineered to meet p95 ≤ 900 ms via batching, memoization, early exits, and selective decode. Heavy verifiers are rate-limited to ≤1 call per request and ≤15% on a rolling window (incident bursts permitted with receipts). In addition to normal-operation metrics, each route SHOULD publish P50/P90 serve costs under a stress scenario (cache-miss + heavy=40%); §D.10 provides matching simulation overlays to contextualize observed deltas.

### D.6 Adversarial evaluation

Inclusion of poisoning scenarios is required given their demonstrated practicality at scale and small-sample pressure [17, 20]. Attack classes cover prompt injection (including RAG-mediated) [19], citation stuffing, paraphrase laundering, threshold probing, source collusion, and temporal confusion. Small-sample poisoning—attacks effective with a near-constant number of poison samples—motivates issuer caps, manifest discipline, and periodic adversarial canaries. Sampling plans and α-spending for sequential tests MAY be pre-registered; results SHOULD report ASR/Q2S/TTD with confidence intervals.

### D.7 Baselines & ablations (locked)

Baselines: (B1) RAG+calibration+human-handoff; (B2) RAG+guardrails (no receipts). Confident-error@τ with temp-scaled confidence; τ fixed in prereg; report absolute/relative deltas with 95% CIs. Ablate each component (manifests, receipts, MSES, heavy) vs both baselines.

### D.8 Helpful-abstention UX

Abstentions must explain why and what next (rephrase, specify jurisdiction, escalate) and expose conflicting evidence. Measure abandonment Δ and time-to-resolution with CIs; show next-step guidance in-receipt. Use a 1–5 rubric (≥4=helpful), A/B test abandonment/satisfaction/time-to-resolution, and log handoffs with ETA for accountability.

### D.9 Limits & future work

Coverage remains the chief limitation: verifier diversity is insufficient to break correlated failure modes, and non-neural specification checks are too narrow to catch subtle policy or schema drift. Priority directions include neuro-symbolic hybrids that bind formal constraints to learned representations, portability to messy, weakly-cataloged corpora (with robust metadata inference and uncertainty flags), and systematic studies of when governance overhead fails ROI in lower-stakes or rapidly changing domains. The agenda explicitly invites negative-result publication—cases where abstention frustrates users, dependence inflates error budgets, or receipts add cost without measurable risk reduction—so that future designs can tighten assumptions, slim contracts, and target only those settings where attestation materially improves outcomes.

### D.10 Monte Carlo simulator: correlation, ranges, validation, power overlays

D.10 Monte Carlo simulator (pointer). A full specification of the simulator—models (beta–binomial; Gaussian/t/vine copulas) [6], [7], inputs (pair FP/FN, $N_f$, $N_c$, cascade rules), validation checks, and outputs (FWER/FDR post-Holm/BY, $m_{\text{eff}}$with CIs, n per arm, and stress-cost overlays)—is provided in Appendix §D. §G.3 draws its sample-size tables from D.1, and Appendix F uses the stress overlays (e.g., heavy=40%, no cache) when reporting P50/P90 serve costs.

*

## CONCLUSION

This proposal began as an exploration of whether governance primitives from earlier computing eras—contracts from SHRDLU's blocks world [33], trails from Memex's hypertext vision [29], receipts from Xanadu's transclusion dream [32]—could be adapted to constrain modern LLM generation. What emerged is not a revival of those historical systems' properties or scope, but a contemporary architecture that borrows their organizing metaphors to structure a verifiable decision-making pipeline: policy-bound gates that decide what may be attempted, fragment-level provenance that anchors what was used, and signed artifacts that attest what was checked. The historical labels are pedagogical wayfinding, not claims of inheritance; the substance is cryptographic proofs, statistical risk budgets, and fail-closed execution contracts applied to a generation stack.

Across this document the limits are stated plainly: no guarantee of source truth, no mechanistic interpretability, no elimination of adversarial risk, no universal applicability, and no production-grade validation to date. The headline targets (≥20% error reduction, ≥70% helpful-abstain, p95 ≤ 900 ms, ≤2.2× cost) are engineering projections rather than measurements; a pilot may show they are unreachable or only attainable at unacceptable trade-offs. To keep claims credible, hypotheses are pre-registered (optional), NO-GO gates are fixed in advance, and negative results are committed for publication. Success would therefore clear a bar set high enough to matter; failure would still yield a rigorous, decision-useful record of what did not work and why.

The ultimate contribution of this work may not be the specific architecture but the demonstration that governed AI systems can and should be evaluated with the same standards applied in clinical trials (pre-specified endpoints, prospective registration, transparent reporting) [4, 37, 39], in safety-critical engineering (hazard analysis, verification & validation, fail-safe design) [39, 40], and in randomized policy experiments (adequate statistical power [5], preregistered analysis plans, credible causal inference) [41]. These principles—pre-registration (optional), sufficient power, independent oversight, and honest reporting of harms and costs—are explicitly adopted here (see §G).

The proposal is now complete. The next step is empirical validation. Scrutiny, replication, and extension are encouraged. Transparent reporting—positive or negative—will be provided so that subsequent work proceeds on evidence rather than aspiration.

## ACKNOWLEDGEMENTS

The author thanks the teams behind the large language models (LLMs) and open-access toolchains used for drafting, translating and editing, and acknowledges the broader open-access AI ecosystem that made this work possible. This paper draws inspiration from three historical systems that foregrounded explainable structure, auditability, and verifiability: Terry Winograd's SHRDLU (explicit state and operations), Vannevar Bush's Memex (preserved research trails), and Ted Nelson's Xanadu (transclusive provenance). My thanks as well to the authors of Anthropic's small-sample poisoning study for empirically sharpening adversarial risks, among those that this "Policy-Governed RAG with Cryptographic Receipts" aims to address.

*

## APPENDICES

### Appendix A — Historical context (mnemonics)

SHRDLU, Memex, and Xanadu are used strictly as mnemonic handles rather than endorsements of their scope or feasibility. Originally, IA revives and operationalizes these ideas as SHRDLU-2025 (situated reasoning + state logs) [33, 34], Memex-2025 (replayable research trails) [29, 30, 31], and Xanadu-2025 (fine-grained, versioned transclusion) [32]. This is in the same spirit as recent reconstruction work in other subfields—for example, the Pucci-by-AI reimplementation that re-animates a century-old rule-and-trace paradigm for machine translation [35]. Here the revival is explicitly operational and bound to three governance primitives—contracts, trails, receipts—without importing the brittleness, incompleteness, or unrealized ambitions of the originals; these labels function purely as pedagogical wayfinding.

The contemporary architecture described here borrows organizing metaphors but relies on modern cryptographic proofs (Merkle trees [21], transparency logs [22]) and portable signed objects (COSE receipts [24]), statistical risk controls (Holm [1]; Benjamini–Yekutieli [2]), and falsifiable evaluation protocols (pre-registered pilots (optional) with NO-GO gates; see §10 and Appendix G).

## Appendix B — Challenge-set governance

Role separation and auditability are hard gates: if unmet, the challenge set is invalid for confirmatory evaluation.

Challenge sets are to be authored under independent authorship: challenge authors and verifier developers SHOULD be disjoint teams (no personnel overlap or code contribution within a 90-day cooling window), and conflict-of-interest disclosures SHOULD be recorded prior to work. Item curators (red-teamers and domain practitioners) operate without access to verifier internals (weights, prompts, thresholds); verifier teams receive only the public schema and historical versions after the evaluation window. A held-out set unsealing protocol SHOULD be followed (two-person rule; hash-pinned manifest; timestamped release). All access SHOULD be audit-logged (who/when/what/why; immutable log; entry hash), and any export SHOULD carry a reproducible digest.

Sizing is power-based per attack class (typically ≥800–1,200 attempts), for a total $n \geq 2{,}000$ pairs, split 50/50 tuning vs held-out evaluation, with 20% reserved for drift checks. The adversarial mix targets (±5 pp) are: Negation 20%, Modality 15%, Paraphrase laundering 20%, Numeric 15%, Temporal 15%, Multilingual 10%, Source-collusion 5%. Each item is double-rated by practitioners with adjudication; labels are entail / neutral / contradict with rationales, using the three-way entailment schema from FEVER [16] and the calibrated, guideline-driven annotation protocol from TRUE [15]; inter-rater $\kappa$ is reported. Sets SHOULD refresh annually (≥30% new) with emergency updates on major policy shifts; prior versions are archived for regression with version hashes. For openness, schemas are published, ≥10% of items are released with summary metrics, and external red teams/bounties are invited—eschewing security-by-obscurity.

## Appendix C — Route demotion protocol

A governed route is automatically demoted to Assistive (under review) when dependence collapses ($m_{eff} < 1.3$ with 95% CI LB < 1.1), when helpful-abstain falls below 60%, or when the red-team pass rate exceeds the pre-registered (optional) threshold. During demotion, the UI displays a warning banner, the route emits no receipts, and outputs are labeled advisory only. Operations initiate root-cause analysis within 5 business days, with a remediation or permanent demotion decision within 30 days, and a public incident note (redacted as needed) within 7 days. Remediation may involve retraining verifiers, tightening gates, or diversifying manifests; changes are logged and versioned. Re-promotion requires a fresh challenge set, re-estimation of $m_{eff}$, and a brief A/B sanity check to confirm restored targets.

## Appendix D — Red-team methods

Scope & governance. Red-team evaluation is conducted on a frozen, hash-pinned snapshot (models, prompts, policies, thresholds, manifests). Role separation is mandatory: red-team authors and verifier developers SHOULD be disjoint personnel and code contributors (90-day cooling-off). Test sets are sealed (hash-committed, time-stamped); unsealing follows a two-person rule with full access logs. Statistical procedures and decision thresholds are pre-registered (optional) (see §G.2).

Attack families & sampling. Six families are exercised: prompt injection (incl. RAG-mediated), citation stuffing, paraphrase laundering, threshold probing, source collusion, and temporal confusion. Per-class sample sizes are power-based (typically 800–1,200 attempts/class), stratified by domain/difficulty. Attack generation mixes scripted agents, internal adversaries, and domain practitioners against the frozen system and strong baselines (RAG+calibration+human-handoff; RAG+guardrails).

Execution discipline. Live-ops telemetry is enabled (guard-railed) to measure Time-to-Detect (TTD) and trigger quarantine/rollback per incident policy. Benign traffic is interleaved to measure latency impact and false blocks. Every attempt is immutable-logged.

Metrics & control. Primary metrics: ASR (attack success rate), Q2S (queries-to-success), TTD (p95 minutes), and benign p95 latency. Multiplicity is controlled as preregistered: Holm (FWER) for contradiction-type tests; Benjamini–Yekutieli (FDR) for support-type rates. Dependence reporting (m_eff) is estimated via beta–binomial overdispersion and copula overlays (Appendix D.1), with 95% CIs. CIs use Wilson or BCa bootstrap as specified in §G.2.e.

Pass/Fail gates (SHOULD meet all vs strongest baseline, 95% CI). ASR ↓ ≥30%, Q2S ≥3×, TTD p95 ≤ 15 min, and benign interactive p95 ≤ 900 ms. Failure on any gate triggers NO-GO for the affected route class until remediation and re-test.

Labeling & adjudication. Outcomes are double-rated by practitioners with adjudication on ties; inter-rater κ is reported. Label schemas and rubrics are versioned and hash-pinned. Any deviation from preregistered procedures is recorded in a deviation log.

Sequential testing. A group-sequential α-spending plan (pre-declared in §G.2.e) MAY stop underperforming classes early or expand promising ones; spending and looks are disclosed in the report.

Logging & receipts. Every attack, detection, mitigation, and decision SHOULD be receipt-logged (input hash; route/policy versions; evidence checks; rationale; timestamps). A public summary (with sensitive fragments redacted) SHOULD include threat matrices, cohort stats, confusion patterns, incident notes, and a released 10% sample of attacks with schemas.

### D.1 Monte Carlo simulator (detailed)

The notebook monte_carlo_error_prop.ipynb simulates claim/answer risk under dependent verifier errors and exports CSV overlays for planning (Appendix G/F). Inputs: pair-level FP/FN ranges; fragments/claim (N_f); claims/answer (N_c); cascade rules; copulas [6], [7] (Gaussian; t with $\nu \in \{4,6,10\}$; vine) and a beta–binomial overdispersion option; plus bootstrap seeding from challenge-set confusion patterns. The simulator propagates pair→claim→answer, reports post-Holm/BY FWER/FDR, m_eff (with CIs), and n per arm required to detect a 20% relative drop at 80–90% power. Outputs include power curves, κ/m_eff sensitivity, and latency/cost stress scenarios (e.g., heavy=40%, no cache). Validation checks: recover independent-test baselines as $\rho \to 0$; reproduce analytic bounds for simple cascades; cross-check low-N with exact binomial. All seeds/configs are logged for replay; a CLI wrapper exports tables referenced in §G.3 (sample size) and Appendix F (ROI).

## Appendix E — Helpful-abstain evaluation

Abstentions are evaluated with a practitioner-facing instrument on a 1–5 Likert scale. Raters judge whether the abstention makes its rationale clear, proposes a concrete next step (e.g., rephrase, escalate, specify jurisdiction/date), and surfaces conflicting or missing evidence. Each route must accrue at least 100 distinct abstentions, each rated by at least three domain practitioners under blinded conditions. The primary metric is the share of items rated ≥4/5, reported with 95% confidence intervals; the pass threshold is a median helpful-abstain of at least 70%. In parallel, task abandonment is measured relative to a control and flagged when the increase exceeds +10% with confidence intervals reported. Results inform route demotion/remediation and guide UX iterations.

## Appendix F — Economic model & calculator

This appendix provides a risk-adjusted ROI calculator and decision gates. The domain-fit gate (F.0) sets a lower bound on error cost: when the estimated loss per incident L is below €5,000, confirmatory efficacy is skipped in favor of ops-only validation (latency/cost stability, receipt integrity) and publication of negative ROI curves. The domain selector (F.3) accepts inputs—including L, abstention cost A, traffic volume and mix, serve cost and latency penalties, routing share r, human review cost C_H, queue/SLA penalties, deferral costs, and security/legal overheads (HSM operations, key ceremonies, legal review, external audits, proof I/O)—and returns the breakeven reduction in confident-error, sensitivity curves, P50/P90 total costs, and a stress scenario (no cache, heavy=40%). A GO recommendation applies when the breakeven Δconf-err is ≤25% at P50 overhead ≤2.2×; otherwise the guidance is NO-GO for efficacy with D1 ops-only validation.

## Appendix G — Pilot roadmap & pre-registration (optional) (two deliverables)

### G.0 Purpose & Scope

Scope. This appendix defines the *pilot process and evaluation only*: deliverables (D1 ops, D2 efficacy), preregistration package, power/sample-size planning, conditional-power policy, timeline & checkpoints, GO/NO-GO rubric, and reporting.

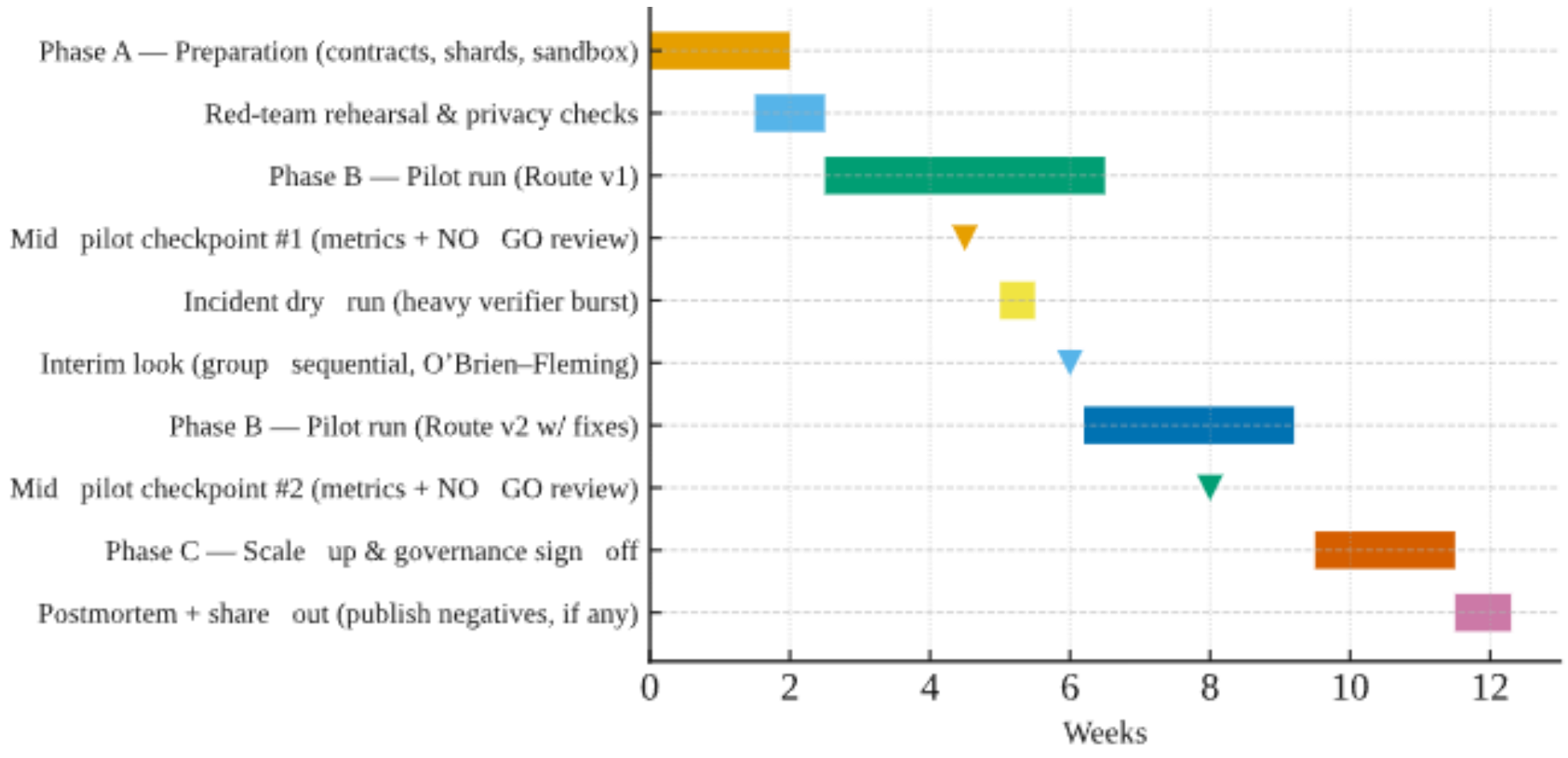

**Fig 7**. Pilot timeline (Gantt with checkpoints)

Substrate selection and revocation alignment are out of scope here—operators shall choose and document per Appendix I and cite the choice in the D1 report.

Cross-refs used herein. Substrates & receipt portability → Appendix I. Economic model/domain-fit gate → Appendix F. Monte Carlo simulator & dependence overlays → Appendix D.10.

### G.1 Deliverables (split: ops vs. efficacy)

#### *D1 — Ops-Only Report (T+90–120 days)*

Goal. Validate operational viability without making efficacy claims.

Inputs. n≈200–400 total queries (shadow + canary; not powered).

Outputs.

1. Latency distribution (p50/p90/p95/p99; with/without heavy verifier).
2. Cost bands (P50/P90 per query: base, caches, HSM, proof I/O, legal review).
3. Abstention analysis (rate; reasons; scope_diagnostics histogram).
4. Receipt integrity (verification pass%, proof size distribution, proof timeout%).
5. Shadow-traffic confident-error estimates (wide CIs; no efficacy claims).
6. Failure taxonomy & incident log (counts; MTTR; mitigations).
7. Operator/UX notes (helpful-abstain feedback; routing pain points).
8. Open-source artifacts (sidecar, schemas, toy verifier, Monte Carlo notebook).
9. Recommendations (domain fit, resource needs for D2, selected substrate per Appendix I).

10. Time box. Ship D1 at T+90–120 days from project start.

### *D2 — Confirmatory Efficacy (when feasible)*

Goal. Test the pre-registered hypotheses with adequate power.

Design. Randomized A/B (query- or session-level) against two strong baselines—(A) RAG+calibration+human-handoff and (B) RAG+guardrails—tuned on the same snapshot; no exposure to test queries; pre-registered stratification.

Primary endpoint (confident-error@τ). Proportion of incorrect emitted answers with calibrated confidence ≥ τ. Pre-register calibration method, τ-selection (frozen on a held-out validation set), and abstention handling (primary excludes abstains; sensitivity treats them as "neither wrong nor right") per §G.2.c; specify analysis methods per §G.2.e and power assumptions per §G.3.

Secondary endpoints. Q2S, helpful-abstain (median ≥4/5), task abandonment, p95 latency, serve cost.

Sample size & power. 80–90% power to detect a 20% relative drop (α=0.05, multiplicity-controlled). See §G.3.

NO-GO gates (D2 only). As specified (effectiveness, usefulness, latency/cost, m_eff).

Analysis sets & deviations. ITT primary; per-protocol sensitivity. Any changes follow the α-spending / amendment rule in §G.2.e.

## G.2 Pre-Registration (optional) Package (OSF)

To be registered on OSF before any D2 data collection (ideally before D1).

### *G.2.a Hypotheses & Endpoints*

Primary/secondary hypotheses; endpoint definitions; metric names; τ targets.

### *G.2.b Design & Randomization*

Arms, randomization unit (query/session), stratification factors, inclusion/exclusion criteria; snapshot discipline (no leakage from test to tuning).

### *G.2.c Calibration, τ, and Abstentions (cross-referenced by D2 primary endpoint)*

(i) Calibration method (e.g., temperature scaling on a held-out set),

(ii) τ selection rule (frozen on the validation set; not re-fit on test),

(iii) Abstention handling: primary analysis excludes abstains; sensitivity analysis treats abstains as "neither wrong nor right" (report both).

### *G.2.d Sample-Size Plan*

Point estimate and sensitivity range; link to power curves; assumptions on baseline error and effect size (see §G.3).

### *G.2.e Analysis Plan & Multiplicity*

Holm (FWER) for contradictions; Benjamini–Yekutieli (FDR) for supports; estimator definitions; CI methods; dependence reporting (m_eff).

### *G.2.f Interim Looks & α-Spending*

Schedule; conditional-power rule (see §G.4); stopping/extension criteria. Interim looks are descriptive only or, if formal, use pre-declared α-spending (O'Brien-Fleming) [4]. Conditional-power triggers do not change α.

### *G.2.g Deviation Log & Amendments*

Template for recording any D1-informed tuning and any threshold changes; amendment rule for what can/can't be changed post-registration.

## G.3 Sample-Size & Power (confirmatory)

Default power target. 85% (range 80–90%), α=0.05 (two-sided), multiplicity-adjusted.

Planning example. Baseline confident-error@τ $p_0$ = 0.15; target relative reduction 20% → $p_1$ = 0.12 (Δ = 3 pp). Required n per arm ≈ 1,543 (Wald with continuity correction; confirmed via Monte Carlo overlays and G*Power exact tests [5]). Safety margin +10% → 1,700 per arm (≈ 3,400 total).

Sensitivity. If $p_0$=0.30 → n per arm≈920; if $p_0$=0.10 → n per arm≈3,166. Inflate n per arm under higher multiplicity or dependence per Monte Carlo overlays (see Appendix D.10).

Reporting. Pre-register the exact n, assumptions, effect metric, and power curves.

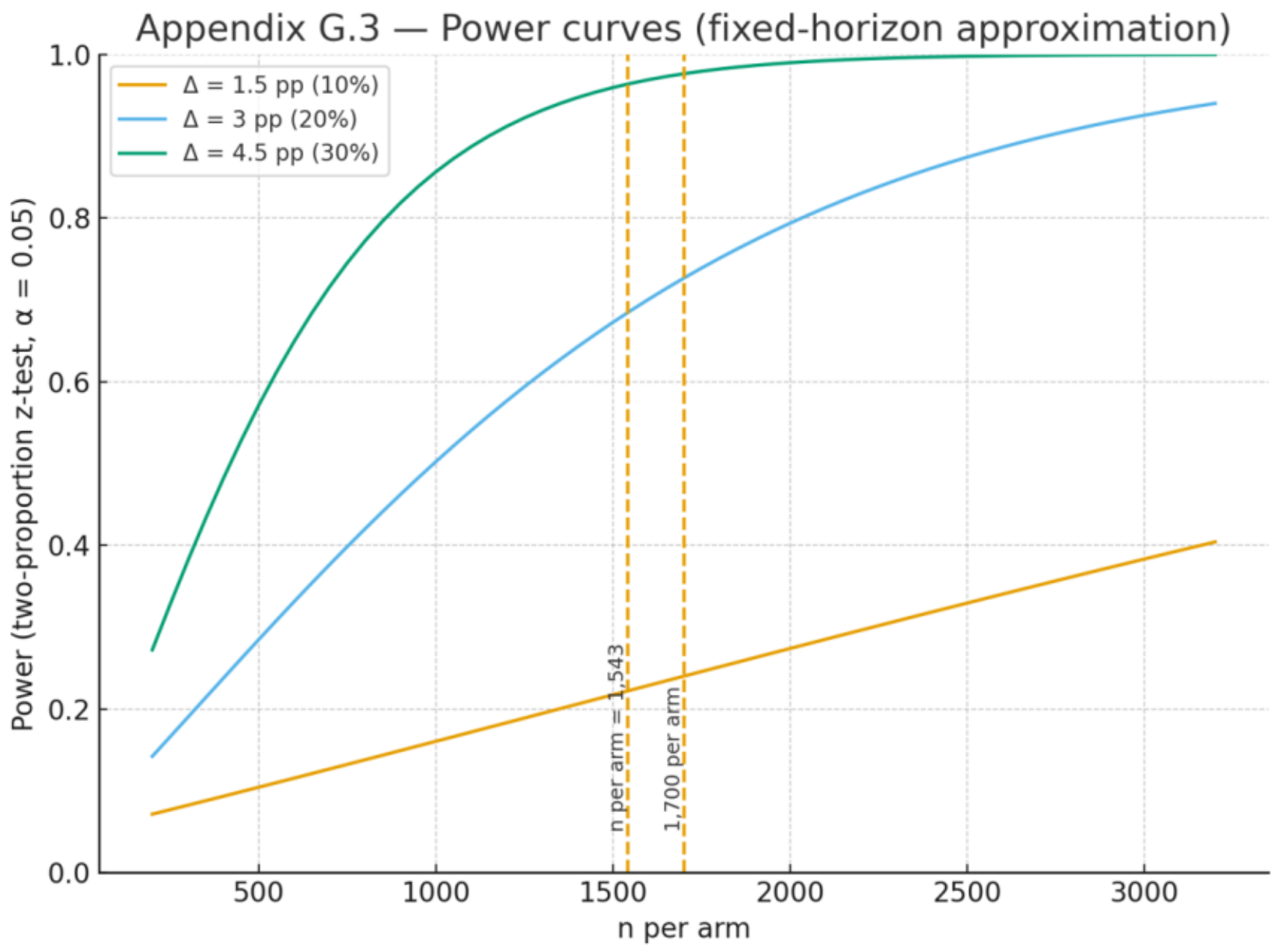

Appendix G.3 – Power curves (fixed-horizon approximation)

### G.4 Conditional-Power Trigger (heavy verifier tie-in)

Definition (conceptual).

low_conditional_power() := True if interim conditional power to detect the pre-registered effect at current n_enrolled falls below 70%.

Operationalization.

- Compute conditional power at each interim look (monthly or every ~300 queries) using observed effect size/variance; honor multiplicity plan.
- If low_conditional_power() is True, enable heavy verifier for disagreement/margin cases and a random 10–20% sample, capped at ≤1 heavy call/request, while meeting p95 ≤ 900 ms.
- Log all trigger events and their effect on BY/Holm budgets; disable when conditional power recovers ≥70%.
- Note. Adaptive information-gathering policy; hypotheses/effect size remain unchanged; pre-register this rule.

### G.5 D1 Deliverable Template (Ops Report)

1. System configuration (routes, thresholds, verifier versions, selected substrate per Appendix I).
2. Latency (p50/p90/p95/p99; with/without heavy; by route).
3. Cost (P50/P90 per query; breakdown: base, caches, HSM, proof I/O, legal). Stress test: P50/P90 serve cost with no cache and heavy=40% (same breakdown); note any SLO breaches and mitigation.
4. Abstentions (rate; top reasons; helpful-abstain ratings distribution).
5. Receipts (verification pass%; proof sizes; timeout rates; disclosure modes).
6. Shadow error estimates (double-adjudicated samples; wide CIs; no efficacy claims).
7. Incidents (counts; MTTR; quarantine/rollback; key events).
8. Operator feedback (deployment toil; UX pain points; training).
9. Open-source drop (sidecar, schemas, toy verifier, Monte Carlo notebook).
10. Recommendations (domain fit, D2 feasibility, partner resourcing).

### G.6 Timeline & Checkpoints

- M0: OSF prereg (D2); D1 deviation log scaffolded.
- M1–M3/4: Build Phase A (Core) + run D1; ship Ops Report T+90–120.
- M4–M6: Stabilize routes; finalize substrate choice (Appendix I); lock thresholds.
- M6 checkpoint: m_eff; latency/cost P50/P90; prelim ASR/Q2S from red-team.
- M8 checkpoint: Component ablations; conditional-power estimate; adjust heavy triggers.
- M10 checkpoint: Freeze thresholds for D2; confirm n per arm feasibility.
- M12: If n per arm unmet → NO-GO for efficacy (publish D1 + recommendation).

- M12–M18+: If feasible, run D2; analyze per prereg; publish results (incl. negative).

### G.7 GO / NO-GO Rubric (concise)

- GO (Efficacy/D2): L ≥ €50k; ≥150 practitioner-hours labeling; legal pre-approval of internal-only receipts; supply ≥ n per arm within 12 months.
- GO (Ops-only/D1): L < €5k or insufficient labeling/legal capacity → run D1; publish ops metrics + shadow estimates; no efficacy claims.
- NO-GO: Cannot deliver D1 under internal-only receipts; legal blocks even hashed fragments; or partner volume cannot reach n per arm within 12 months.

### G.8 Partner & Domain Criteria

- Domain fit. Use Appendix F calculator. If L < €5k, efficacy not pursued; D1 only.
- Volume. Throughput sufficient to reach n per arm in ≤12 months (see G.3).
- Labeling. ≥150 practitioner-hours (or staged; adjudication on disagreements).
- Legal. Pre-approve internal-only receipts (COSE/JOSE; hashes+selectors; on-prem HSM). Redaction default hash/digest for PII/PHI; full text only for public regs or partner-approved corpora.
- Abstention tolerance. Accept 20–35% abstentions if helpful-abstain ≥70% and abandonment ≤ +10% vs control.

### G.9 Legal & Receipt Modes

- Default. Internal-only receipts (hash/digest; selectors; on-prem HSM); no public transparency log (Phase C deferred).
- Portability. Receipts are COSE/JOSE with Merkle multiproofs; verifiable offline.
- Revocation. Follow KRN; mirror any substrate-specific revocations (Appendix I) and reject receipts tied to revoked keys unless re-attested.
- Disclosure policy fields. disclosure_scope: internal-only|partner|public; fragment_mode: hash|digest|full.

### G.10 Data, Ethics & Privacy

PII/PHI scanning before receipt finalization; mask or abstain (privacy reason). Challenge-set governance per Appendix B (independent authorship, adversarial mix, κ reporting). Publish D1 ops report and D2 results (positive or negative); release schemas and toy verifier.

### G.11 Reporting & Publication

- D1: Public ops report within 120 days, minus confidential fragments.
- D2: Full analysis per prereg (Holm/BY, dependence reporting, CI bands), plus ablations vs both baselines.
- Negative results: If NO-GO is triggered (domain fit, time box, or efficacy), publish rationale and lessons learned.

### G.12 References to Other Appendices

- Appendix D.10: Monte Carlo overlays → sample-size curves under dependence (beta–binomial; copulas; bootstrap).
- Appendix F: Economic calculator; domain-fit gate (L < €5k → ops-only).
- Appendix I: Substrates & trade-offs (HSM sidecar, CT/Rekor, ICP/IC-Guard); procurement checklist.

### G.13 (Optional) Worked Sample-Size Note (paste into OSF prereg)

Assumptions: $p_0=0.15$, target 20% relative reduction → $p_1=0.12$; $\alpha=0.05$ two-sided; power=85%; Holm/BY multiplicity accounted for via simulation overlays → n per arm = 1,543; safety margin +10% → 1,700 per arm. Sensitivity: if $p_0\in[0.10,0.30]$, n per arm$\in$[3,166, 920].

### G.14 (Optional) Conditional-Power Helper (pseudo-API)

```
def low_conditional_power():

    """

    Returns True if interim conditional power to detect the pre-registered

    20% relative reduction at current enrollment is < 0.70.

    When True, enable heavy verifier for disagreements, margins, and 10–20% random,

    respecting latency p95 ≤ 900 ms and ≤1 heavy call/request.

    """

    n = enrolled_queries()

    eff = interim_effect_estimate()

    return estimate_conditional_power(n, eff, alpha=0.05, target_rel_drop=0.20) < 0.70
```

## Appendix H — Substrate Quick Reference (non-normative)

Purpose. Handy crib sheet for operators; all normative details live in Appendix I.

Use. Choose one of: HSM sidecar • CT/Rekor • certified-data (e.g., ICP/IC-Guard). For comparison and policy requirements, see Appendix I.

Pointer. Decision checklist, revocation alignment, latency/size targets, and procurement steps → Appendix I §§I.3–I.8.

This page is an indicative summary; authoritative guidance is in Appendix I.

| Criterion | HSM Sidecar | CT/Rekor | Certified-Data (e.g., ICP) |
|---|---|---|---|

| Criterion | HSM Sidecar | CT/Rekor | Certified-Data (e.g., ICP) |
|---|---|---|---|
| Portability | ✅ COSE/JOSE | ✅ COSE/JOSE | ✅ COSE/JOSE |
| Revocation | Manual KRN | CT monitors | Protocol KRN → mirror to proposal KRN |
| Latency (p95) | 50–100 ms | 200–400 ms | 100–200 ms (est.) |
| Ops toil | High | Medium | Low/Medium |
| Cost (10M req/mo) | ~€20k | ~€5k | ~€8k (est.) |
| Audit ecosystem | Custom/internal | Strong (OSS/industry) | Emerging |

When to choose which.

HSM Sidecar: strict residency, minimal latency, internal audit only.

CT/Rekor: need public/consortium audit and mature monitoring.

Certified-Data: you already depend on the protocol, or threshold signatures/certified reads simplify ops.

## Appendix I — Implementation Substrates & Trade-offs (non-normative, canonical)

Status (informative). Appendix I centralizes substrate guidance; the spec remains substrate-agnostic. The only normative rules here are portable COSE/JOSE receipts with offline verification and revocation alignment via KRN mirroring; everything else is guidance.
Cross-refs. Cited from §§8–9 (manifests/receipts), Appendix G (pilot), and Appendix H (quick reference).

### I.0 Purpose & Scope

This appendix catalogs options for anchoring manifest roots and signing receipts. The spec remains substrate-agnostic: the choice of substrate must not change receipt format or verifier logic.

### I.1 Normative rules (the only MUSTs)

Portable receipts. All receipts SHOULD be COSE/JOSE objects that carry the answer hash/metadata, Merkle multiproofs, signer KID/KRN, and disclosure policy, and are verifiable offline by a standalone verifier (no live substrate calls). Substrate choice SHOULD NOT alter receipt fields or verification logic.

Revocation alignment. If a substrate publishes revocations/rotations, those SHOULD be mirrored into the proposal's KRN on an SLO of p95 ≤ 10 min / p99 ≤ 30 min; verifiers check both channels and fail-closed on mismatch/lag, with auditable logs.

Everything else in this appendix is informative guidance to help pick and run a substrate; it does not add new MUSTs.

## I.2 Substrate options (informative)

### *A. HSM-Signed Sidecar (on-prem or VPC)*

Private append-only log (e.g., object storage + integrity checks), org keys in HSMs; optional RFC-3161 notarization. Pros: control, low latency, residency. Cons: ops toil; public audit needs extra steps.

### *B. Transparency Logs (CT/Rekor/Sigstore-like)*

Mature tooling and public verifiability; disclosure policy required; proof sizes grow with log depth.

### *C. Certified-Data Protocols (e.g., ICP/IC-Guard)*

Threshold-ECDSA signatures and certified reads; platform coupling and cost model trade-offs; emerging audit ecosystem.

## I.3 Decision checklist (copy/paste; informative)

- Portability (MAY): Offline-verifiable COSE/JOSE receipts?
- Revocation (MAY): KRN mirror wired; dual-check enforced; fail-closed verified.
- Latency (SHOULD): Added p95 ≤ 200 ms on governed routes.
- Ops model (SHOULD): Key ceremonies, rotations, monitors, incident runbook.
- Cost (SHOULD): Include substrate costs in P50/P90 serve cost.
- Disclosure (MAY): Internal-only mode supported (hash/digest; no public fragments) unless legal sign-off.

## I.4 Indicative comparison (informative; validate in D1)

Sidecar ≈ 50–100 ms p95; CT/Rekor ≈ 200–400 ms; certified-data ≈ 100–200 ms; all retain COSE/JOSE portability. Treat figures as targets to confirm in D1 Ops-Only Report.

## I.5 Security & governance notes (informative)

Two-person admin, HSM-backed keys, scheduled rotation; transparent KIDs in receipts; auditable key use. On mirror lag/mismatch, REVOKED-PENDING-MIRROR and operator alert; quarantine + re-attest on new keys per incident runbook.

### I.6 Performance & proof SLO pointers (informative)

Governed routes aim for p95 ≤ 900 ms E2E; proof SLOs: proof_size_p90 < 64 KB, proof_verify_p95 ≤ 200 ms; degrade to Lite Receipt or ABSTAIN on proof timeout. (Authoritative SLOs live in core sections and Appendix J)

### I.7 Receipt modes & disclosure (informative)

Default internal-only receipts (hash/digest selectors; on-prem HSM). Partner/public modes require legal/DPO sign-off; receipts record disclosure_scope and fragment_mode.

### I.8 Procurement & migration (informative)

Procurement demo set: offline verification; revocation mirror drill; latency test; ops runbook; cost sheet; legal memo on disclosure.

Migration path: Start with HSM sidecar in Phase A/B; if public verifiability is needed later, move to CT/Rekor or a certified-data substrate without changing receipt format.

## Appendix J — Operational defaults & SLO tables

Purpose. This appendix defines route-level operational defaults (latency budgets, timeouts, retries) and verifiable SLOs. Routes MAY tighten these defaults but SHOULD NOT exceed them without logging an SLO change in the route manifest.

### J.1 Latency budget (per governed route)

| Segment | Default p95 budget | Notes |
|---|---|---|
| Retrieval | 250 ms | Includes ANN/reranker if used in “cheap/small” tiers. |
| Cheap verifiers | 120 ms | Fast heuristics / filters. |
| Small verifiers | 180 ms | Second-pass NLI/reranker. |
| Heavy verifiers (if invoked) | 220 ms | Only when routed; otherwise reallocated to other segments. |
| Proof assembly (Merkle multiproofs) | 80 ms | Bounded by proof SLOs in J.3. |
| Signing | 50 ms | HSM or software signer. |
| End-to-end target | ≤ 900 ms (p95) | Must be met under governed conditions. |

Note: If heavy verifiers are skipped, their budget is not “spent” and E2E p95 still ≤ 900 ms.

### J.2 Timeouts, retries, and backoff

| Operation | Timeout (ms) | Retries | Backoff | Fail-mode |
|---|---|---|---|---|
| Retrieval RPC | 250 | 1 | Exponential (×2), jitter | ABSTAIN on exhaustion |
| Cheap verifier | 60 | 1 | Fixed 30 ms | Skip to small; log |
| Small verifier | 120 | 1 | Exponential (×2), jitter | Route to heavy or ABSTAIN per policy band |
| Heavy verifier | 220 | 0 | — | ABSTAIN if not completed |
| Proof build | 300 | 0 | — | Degrade to Lite Receipt or ABSTAIN('proof_timeout') |
| Signing (HSM) | 120 | 1 | Fixed 60 ms | ABSTAIN if failure |
| KRN mirror fetch | 300 | 1 | Fixed 150 ms | Fail-closed (REVOKED-PENDING-MIRROR) |

### J.3 Proof SLOs (Merkle multiproofs)

| Metric | SLO | Measured at |
|---|---|---|
| proof_size_p90 | < 64 KB | Route reference hardware |
| proof_size_p99 | < 256 KB | Route reference hardware |
| proof_verify_p95 | ≤ 200 ms | Standalone verifier |
| proof_timeout_ms (per request) | 300 | Route configuration |

### J.4 Revocation, mirrors, and drift detection

| Control | Default | Action on breach |
|---|---|---|
| KRN mirror freshness SLO | ≤ 5 min behind substrate | Mark REVOKED-PENDING-MIRROR; fail-closed |
| Revocation check | Dual-channel (substrate + local KRN) | Reject if either fails |
| Reconciliation job | Every 15 min | Backfill gaps; re-attest receipts where feasible |
| Canary/drift monitors | Rolling on fresh traffic | Auto-quarantine route/shard on trigger (see §7.4) |

### J.5 Throughput & rate limits (anti-extraction)

| Limit | Default | Notes |
|---|---|---|
| Queries per route (burst) | 50 QPS | Token-bucket |
| Queries per org (sustained) | 10 QPS | Across routes |
| Receipt disclosure mode | Lite by default | Full only with DPO/Legal sign-off |
| Selector length cap | 128 bytes | To reduce linkability |
| Evidence redaction | Enabled | PII/PHI scan pre-finalization |

### J.6 Change control & reporting

SLO changes SHOULD be recorded in the route manifest with: author, timestamp, rationale, before/after values, and snapshot IDs (SHA-256).

Monthly report SHOULD include: E2E p95, segment p95s, proof SLOs, ABSTAIN rates, confident-error@τ, and any REVOKED-PENDING-MIRROR incidents.